\documentclass[aps,twocolumn,showpacs,preprintnumbers,nofootinbib,prd,superscriptaddress,groupedaddress,10pt]{revtex4-1}
\usepackage{graphicx,amssymb,amsmath,amsthm,amsfonts,epsfig,epsf}
\usepackage[usenames]{color}
\usepackage{epstopdf}

\usepackage{aas_macros}
\usepackage{bm}
\usepackage{dcolumn}
\usepackage[latin1]{inputenc}
\usepackage{latexsym}
\usepackage{rotating}
\usepackage{longtable}

\usepackage{enumerate}
\usepackage{tensor,multirow}
\usepackage{url}
\usepackage[linktocpage]{hyperref}
\usepackage{graphicx,amssymb,amsmath,amsthm,amsfonts,epsfig,times,bm}
\usepackage{aas_macros}
\usepackage{cleveref}
\newcolumntype{C}[1]{>{\centering\arraybackslash}p{#1}}

\usepackage{amsmath,amssymb}
\usepackage{tensor}

\newcommand{\tn}[1]{{\rm #1}}

\def\be{\begin{equation}}
\def\ee{\end{equation}}

\newcommand{\bea}{\begin{eqnarray}}
\newcommand{\eea}{\end{eqnarray}}

\def\nn{\nonumber}

\begin{document}
\title{Constraining Black Holes with Light Boson Hair and Boson Stars using Epicyclic Frequencies and QPOs}

\author{
Nicola Franchini,$^{1,2}$ 
Paolo Pani,$^{1,3}$
Andrea Maselli,$^4$
Leonardo Gualtieri,$^{1}$
Carlos~A~R~Herdeiro,$^{5}$
Eugen Radu,$^{5}$
Valeria Ferrari$^{1}$
}

\affiliation{$^{1}$ Dipartimento di Fisica, ``Sapienza'' Universit\`a di Roma \& Sezione INFN Roma1, Piazzale Aldo Moro 5, 00185, Roma, Italy}
\affiliation{$^{2}$ School of Mathematical Sciences, University of Nottingham, University Park, Nottingham, NG7 2RD, UK}

\affiliation{${^3}$ CENTRA, Departamento de F\'{\i}sica, Instituto Superior T\'ecnico -- IST, Universidade de Lisboa -- UL, Avenida Rovisco Pais 1, 1049 Lisboa, Portugal}
\affiliation{$^{4}$ Theoretical Astrophysics, Eberhard Karls University of Tuebingen, Tuebingen 72076, Germany}
\affiliation{${^5}$ Departamento de F\'{\i}sica da Universidade de Aveiro and Center for Research and Development in Mathematics and Applications (CIDMA), Campus de Santiago, 3810-183 Aveiro, Portugal.}
\begin{abstract}
Light bosonic fields are ubiquitous in extensions of the Standard Model. Even when minimally coupled to gravity, these fields might evade the assumptions of the black-hole no-hair theorems and give rise to spinning black holes which can be drastically different from the Kerr metric. Furthermore, they allow for self-gravitating compact solitons, known as (scalar or Proca) boson stars.
The quasi-periodic oscillations (QPOs) observed in the X-ray flux emitted by
accreting compact objects carry information about the strong-field region, thus providing a powerful tool to constrain deviations from Kerr's geometry and to search for exotic compact objects.
By using the relativistic precession model as a proxy to interpret the QPOs in terms of geodesic frequencies, we investigate how the QPO frequencies could be used to test the no-hair theorem and the existence of light bosonic fields near accreting compact objects. We show that a detection of two QPO triplets
with current sensitivity can already constrain these models, and that the future eXTP mission or a LOFT-like mission can set very stringent constraints on black holes with bosonic hair and  on (scalar or Proca) boson stars. The peculiar geodesic structure of compact scalar/Proca boson stars implies that these objects can easily be ruled out as alternative models for X-ray source GRO J1655-40.
\end{abstract}

\maketitle


\section{Introduction}\label{intro}

As the simplest macroscopic objects in the universe, black holes (BHs) are unique probes of fundamental physics and of gravity in extreme conditions~\cite{Berti:2015itd}. Stationary vacuum solutions of Einstein's equations are uniquely described by the Kerr metric~\cite{Kerr:1963ud}, which depends only on two parameters, namely the mass $M$ and the angular momentum  $J:=a^\star GM^2/c$ - cf., $e.g.$~\cite{Chrusciel:2012jk}. BH no-hair theorems guarantee that this uniqueness holds true also for a variety of matter fields minimally coupled to gravity (see~\cite{Robinson} for a historical account and~\cite{Volkov:2016ehx,Herdeiro:2015waa} for recent reviews). These theorems strongly suggest that the plethora of BHs which populate our universe in various mass ranges are universally described by the Kerr family~\cite{Cardoso:2016ryw}.

This paradigm has been recently challenged. The authors of Ref.~\cite{Herdeiro:2014goa} realized that --~within general relativity~-- the no-hair theorem can be evaded if matter fields are time dependent in such a way that their stress-energy tensor remains stationary. The same type of time dependence occurs in the familiar stationary states of non-relativistic quantum mechanics. Even when minimally coupled to gravity, these matter fields give rise to stationary BH solutions which --~in addition to the mass $M$ and angular momentum $J$~-- are characterized by an extra parameter, namely by a Noether charge $Q$ (cf.~Ref.~\cite{Herdeiro:2015gia} for a detailed discussion).
These solutions have been found for (massive, complex) scalar~\cite{Herdeiro:2014goa} and for vector~\cite{Herdeiro:2016tmi} fields and can be dramatically different from the classical Kerr solution.\footnote{Extensions with scalar self-interactions~\cite{Kleihaus:2015iea,Herdeiro:2015tia} and electromagnetic fields~\cite{Delgado:2016jxq} have also been constructed.} A visually striking illustration of these differences is provided by the corresponding BH shadows~\cite{Cunha:2015yba,Vincent:2016sjq}. Interestingly, these hairy BHs are related to the threshold of the superradiant instability of Kerr BHs, which also provides one possible natural mechanism for their formation~\cite{Brito:2014wla,Sanchis-Gual:2015lje}: in the presence of light bosons, Kerr BHs which can form in the gravitational collapse or in a merger can lose angular momentum due to the occurrence of the superradiant instability; part of the energy is transferred to a bosonic condensate and the BH develops a bosonic ``hair'' (cf.~Ref.~\cite{Superradiance} for an overview of BH superradiance).
These solutions interpolate between a Kerr BH (when $Q=0$) and a \emph{boson star} (when $Q$ is maximum). The latter is a self-gravitating compact object made of a scalar~\cite{Kaup:1968zz,Ruffini:1969qy} or a Proca~\cite{Brito:2015pxa} condensate. \textit{Scalar boson stars}, hereafter simply ``boson stars" following widespread terminology,  have been studied for long time (cf.~Refs.~\cite{Schunck:2003kk,Liebling:2012fv} for reviews) as an example of exotic compact objects that can mimic the properties of a BH but lack an event horizon~\cite{Berti:2015itd}. Somewhat surprisingly, their vector cousins, \textit{Proca boson stars}, hereafter simply ``Proca stars" following~\cite{Brito:2015pxa}, have only recently been constructed.

While BHs with bosonic hair and boson/Proca stars exist for any nonvanishing boson mass $m_b=\mu \hbar$, their ADM mass, $M$, is restricted\footnote{We stress that we use $G=c=1$ units, in which $\mu$ has the dimensions of an inverse length. In physical units, the dimensionless quantity reads $G Mm_b/(\hbar c)=Mm_b/(M_{Planck}^2)$. Thus, the ADM mass is restricted by $Mm_b/(M_{Planck}^2)<1$.} to $M\mu\lesssim1$.  This requires an ultralight bosonic field, $m_b\lesssim 10^{-11}(10 M_\odot/M)\,{\rm eV}$. Much higher boson field masses, say of $\sim {\rm GeV}$ can lead to hairy BHs with astrophysical ADM masses $\gtrsim M_\odot$, if bosonic self-interactions are included~\cite{Herdeiro:2015tia}. But, as observed in~\cite{Herdeiro:2015tia}, for the BH \textit{horizon mass} to be $\gtrsim M_\odot$ or higher, an ultralight bosonic field is still required.

Such ultralight fields are ubiquitous in extensions of the Standard Model, such as the hidden $U(1)$ sector~\cite{Jaeckel:2010ni,Essig:2013lka}, and are predicted from string theory models such as the ``string axiverse'' scenario~\cite{Arvanitaki:2009fg}. Furthermore, sub-eV fields are natural dark-matter candidates that are attracting increasing attention, see $e.g.$~\cite{Hui:2016ltb}, in light of the negative results of WIMP-like dark matter searches~\cite{Bertone:2004pz}.
In several models, the mass spectrum of ultralight bosons roughly ranges from $\sim10^{-33}\,{\rm eV}$ to the sub-${\rm eV}$ scale~\cite{Arvanitaki:2009fg}. Interestingly, the superradiant instability time scale depends strongly on the value of $M\mu$ and it is only effective for BHs with mass $M\sim 1/\mu$~\cite{Superradiance}. Thus, depending on the value of the boson mass, Kerr BHs might be unstable in a certain mass range and dynamically evolve~\cite{Brito:2014wla} towards the hairy solutions discussed here, whereas much heavier/lighter Kerr BHs would remain stable. In other words, these models predict that both Kerr BHs and hairy BHs or boson stars can exist in the universe, and it is important to devise tools to distinguish among different solutions.

The spectrum emitted by accreting compact objects provides us with invaluable information to study the properties of strong gravitating systems.
Most of the radiation originates very deep in the gravitational field of these objects and holds the potential to test the near-horizon region of BHs.
Indeed, X-ray observations are a powerful tool to study accreting BHs, either of stellar mass in binary systems or supermassive
in Active Galactic Nuclei (AGNs), and accreting neutron stars. The accretion disks of these objects have a soft X-ray
continuum emission, whose highest temperature allows to measure the innermost disk radius, which is quite naturally assumed to
correspond to the Innermost Stable Circular Orbit (ISCO). It is thus possible to infer the spin of stellar mass
BHs, since the latter is directly related to the ISCO radius, assuming the Kerr model (for a review see~\cite{McClintock:2011}). A powerful
diagnostic technique is based on the analysis of the broad iron K$\alpha$ line, which is observed in
the X-ray spectrum of accreting BHs (both in binaries and in AGNs) around $\sim 6$~keV. The shape of this line carries the imprint of the
emissivity and of the geometrical properties of the inner disk region (for a review see e.g.~\cite{Fabian:2013,Bambi:2015ldr}).

The first proposals to study the geodesic motion in the strong field, inner disc region using the fast variability of
the X-ray flux emitted by matter close to BHs and neutron stars date back to the 1970s (e.g. \cite{Sunyaev:1972}).
However, this approach has only been applied twenty years later, when Quasi Periodic Oscillations (QPOs) were discovered
in the X-ray flux from accreting compact objects. QPOs from accreting BHs in binaries have frequencies up to
$\sim450$~Hz, close to those expected from bound orbits near the ISCO (for a review see~\cite{vanderklis:2006}).
QPOs have also been detected in the X-ray flux from AGNs~\cite{Gierlinski:2008}. 
Several models have been proposed to
describe this phenomenon; almost all of them involve frequencies associated to the orbital motion of matter in the inner disk
region. In this strong-field region, a Newtonian description --~but also a weak-field expansion of general relativity~-- are inadequate.

The first suggestion that QPOs may be used to test the strong-field regime of gravity were based on a simple model in which the QPO frequencies correspond to the geodesic motion of a test particle, e.g.\ a blob of fluid in the accretion disk~\cite{1990ApJ...358..538K,Stella:1998mq}. Although such simplistic model has been ruled out as an origin for QPOs (because it would imply a large number of higher harmonics which are not observed in the spectrum~\cite{2004ApJ...606.1098S} or because the blob of fluid can be quickly destroyed by shear~\cite{2005MNRAS.357.1288B}), viable extensions of the original idea (e.g., considering blobs extended along the azimuthal direction or due to some replenishment mechanism for the blobs) are still under consideration. Furthermore, virtually all models aiming at explaining the QPOs observed from accreting compact objects are based on the motion of fluid along the accretion disk as key ingredient. These include more realistic models which are based on
an analysis of eigenmodes of accretion fluid~\cite{1999PhR...311..259W,2003MNRAS.341..832Z,2004ApJ...617L..45B,2005AN....326..820K,2006CQGra..23.1689A,2015MNRAS.446..240M,2016MNRAS.456.3245M,2015arXiv151007414M} (cf., e.g, Ref.~\cite{Motta:2016vwf} for a recent review on QPOs in BH X-ray binaries).

Due to its connection to the motion of matter near the ISCO, if properly modelled the QPO signal can provide a powerful diagnostic of strong
gravitational fields. Compared to neutron stars, BHs are especially promising in this respect, due to their simplicity, the lack of strong magnetic fields, and the absence of a ``hard'' surface, which can alter considerably the dynamics of matter inflow.
Simultaneous QPO modes in BHs have so far been detected only in a
few cases; the BH X-ray binary GRO J1655-40 is the only BH system in which \emph{three} simultaneous QPOs were observed. 
This triplet corresponds to two high-frequency modes (in the ratio $3:2$, as observed in several other sources) and a low-frequency mode. 
Although high-frequency QPOs from BHs are much weaker and intermittent than their neutron-star
counterparts, their frequency appear to be more constant over the baseline, and hence better suited for precision tests.
All these properties are challenging to explain within a single QPO models. For example, the Relativistic Precession Model (RPM) considered in this work [see Section~\ref{sec:QPO}] can explain the QPO triplet naturally, but does not explain the rational ratio of the twin kHz QPOs. On the other hand, resonance models~\cite{Kluzniak_Abramowicz:2001,Abramowicz_Kluzniak:2001} can explain this fixed ratio in terms of nonlinear resonances between modes of accretion disk oscillations, but require some extension~\cite{Abramowicz:2004je} to accommodate the QPO triplet of GRO J1655-40.
Recently, by interpreting the QPO triplet with the RPM, Ref.~\cite{Motta:2013wga} derived precise measurements of the BH mass and spin, the former being in agreement with the mass obtained from optical observations.

Next-generation large area X-ray instruments --~such as the upcoming eXTP satellite~\cite{eXTP,Zhang:2016ach} and the proposed detector LOFT~\cite{feroci2012large}~-- are expected to detect several simultaneous QPOs in a variety of BHs, and to measure their frequencies to high precision and accuracy, so
that a better modeling of the QPO signal and quantitative tests of gravity in the strong-field
regime will become feasible.
An analysis of the QPO frequencies to test the nature of the Kerr metric was performed in Ref.~\cite{Bambi:2013fea} for a phenomenological deformation of the Kerr metric, whereas 
the QPO spectroscopy of spinning BHs in modified theories of gravity with quadratic curvature terms has been recently performed in Ref.~\cite{Vincent:2013uea} and in Refs.~\cite{Maselli:2014fca,Maselli:2017kic}, finding that --~especially in the case of Einstein-dilaton-Gauss-Bonnet gravity~\cite{Berti:2015itd}~-- interesting constraints can be obtained by future QPO measurements.

In this paper, we calculate the azimuthal and epicyclic frequencies of spinning BHs with scalar and Proca hair, and find that these can differ dramatically from their Kerr counterpart.
This is no surprise, since spinning BHs with scalar and Proca hair are not necessarily described by small deformations of the Kerr solution as in most theories of modified gravity~\cite{Berti:2015itd}. Indeed, these solutions interpolate between Kerr BHs and boson stars, the latter being horizonless objects more akin to stars than to BHs.
Using the RPM, we show that the differences between the
QPO frequencies of a Kerr BH and of a BH with bosonic hair or of a boson star are already detectable with current facilities, and that future large-area X-ray instruments such as eXTP will provide very stringent constraints on these solutions.

Recently, similar tests have been performed on BHs with scalar hair and on (scalar) boson stars using the iron K$\alpha$ line expected in the reflection spectrum of accreting BHs~\cite{Ni:2016rhz,Cao:2016zbh}. Our results using QPO diagnostics are qualitatively similar to those derived with the iron K$\alpha$ line and, in addition, extend those conclusions to the vector case.

\section{Spinning black holes with bosonic hair}\label{sec:hairyBHs}
We stress that the objects under inspection arise as solutions of standard general relativity minimally coupled to a (massive, complex) scalar or vector field.
The theory can be described by the following action (we use units in which $G=c=1$)
\begin{eqnarray}\label{theory}
S&=&\int
d^{4}x\sqrt{-g}\left(\frac{R}{16\pi}-\frac{1}{2}\partial_{a}\bar\Psi\partial^{a}\Psi-\frac{1}{2}\mu_S^2 \bar\Psi\Psi \right.\nn\\
&&\left.
-\frac{1}{4} F_{ab}\bar F^{ab}-\frac{1}{2}\mu_V^2 \bar A_a A^a \right)\,,
\end{eqnarray}
where $R$ is the Ricci curvature, $\Psi$ is the (complex) scalar field, $F_{ab}=\partial_a A_b-\partial_b A_a$, $A_a$ is the (complex) Proca field and a bar denotes complex conjugation. Since in our study we consider the scalar and the vector cases separately, we unify the notation by setting $\mu=\mu_S$ or $\mu=\mu_V $ for the mass term\footnote{Note that, in geometrical units $G=c=1$, $\mu$ has dimensions of an inverse mass, cf.\ Eq.~\eqref{theory}. The physical mass of the field is simply $m_b=\mu\hbar$.} of the scalar field and of the vector field, respectively.
In the action above, we have assumed that the bosonic field is very weakly coupled to ordinary matter (e.g.\ to the plasma of an accretion disk) so that we can effectively ignore this coupling. This is a very natural assumption if the field is a dark-matter candidate and is also consistent with the fact that ultralight fields have not been detected so far in particle-detector experiments~\cite{Jaeckel:2010ni,Essig:2013lka}.

The field equations of the theory are the standard Einstein equations, $G_{ab}=8\pi T_{ab}$, where the stress-energy tensor reads
%
\begin{equation}
 T_{ab}=\partial_{\left(a\right.}\bar{\Psi}\partial_{\left.b\right)}\Psi
-\frac{1}{2}g_{ab}\left[\partial_{c}\bar{\Psi}\partial^c\Psi +\mu^2\bar{\Psi}\Psi\right]  \,,
\end{equation}
for the scalar field, and
%
\begin{eqnarray}
 T_{ab}&=&-{F}_{c\left(a\right.}\bar{{F}}_{\left.b\right)d} g^{cd}
-\frac{1}{4}g_{ab}F_{cd}\bar{{F}}^{cd} \nn\\
&&+\mu^2\left({A}_{\left(a\right.}\bar{{A}}_{\left.b\right)}
- \frac{1}{2}g_{ab}{A}_c\bar{{A}}^c\right)\,,
 \end{eqnarray}
%
for the vector field.
Einstein's equations are supplied by either the Klein-Gordon equation or the Proca equation,
\begin{eqnarray}
 \square\Psi=\mu^2 \Psi \,,
\\
 \nabla_a{F}^{ab}=\mu^2{A}^b\,,
\end{eqnarray}
in the scalar and vector case, respectively.

The Kerr metric is clearly a solution of the above field equations when $\Psi=0$ and $A_a=0$.
However, it is not the only stationary and axisymmetric regular BH in this theory~\cite{Herdeiro:2014goa,Herdeiro:2015gia,Herdeiro:2016tmi}. A new family of solutions describing \textit{Kerr BHs with scalar or vector hair}\footnote{This terminology intends to stress that these solutions are continuously connected to the vacuum Kerr solution, in the same spirit as that used for the Kerr-Newman~\cite{Newman:1965my} or Kerr-Sen~\cite{Sen:1992ua} metrics. Of course, in general the hairy BHs are different from Kerr, like the last two examples, and only reduce to vacuum Kerr in a particular limit. To avoid confusion, in the following we will refer to them simply as hairy BHs.} [hereafter hairy BHs (HBHs)] can be obtained when the scalar or the vector fields have an oscillatory behavior, namely
\begin{eqnarray}
\Psi &=\phi(r,\theta)e^{i(m\varphi-w t)} \,,
\\
 {A}_a &=B_a(r,\theta) e^{i(m\varphi-w t)} \,,
\end{eqnarray}
where $m$ is an integer and $w$ is the characteristic frequency of the oscillation.
In this case the stress-energy tensor is stationary. As a consequence, the metric remains stationary and it can be described by the following ansatz
\begin{eqnarray}
ds^2&=&-e^{2F_0} N dt^2+e^{2F_1}\left(\frac{dr^2}{N }+r^2 d\theta^2\right)\nn\\
&&+e^{2F_2}r^2 \sin^2\theta (d\varphi-W dt)^2\,, \label{metric}
\end{eqnarray}
where $N:= 1-{r_H}/{r}$ and $r_H$ is the BH horizon, while $F_i$ and $W$ are functions of $(r,\theta)$. The metric and the scalar (resp. the vector) field can be obtained by solving numerically Einstein's equations coupled to the Klein-Gordon (resp. Proca) equation~\cite{Herdeiro:2014goa,Herdeiro:2015gia,Herdeiro:2016tmi}.
%
%

The numerical solutions are defined by five parameters: the ADM mass $M$, the
ADM angular momentum $J$, a Noether charge $Q$,  the azimuthal harmonic index $m$ and the node number $n$~\cite{Herdeiro:2014goa,Herdeiro:2015gia}. In this work, we shall restrict to the most interesting cases and consider $m=1$ and only the fundamental states ($n=0$ and $n=1$ for the scalar and vector case, respectively). It is convenient to introduce a normalized Noether charge $q:=mQ/J$, such that $0\le q\le 1$~\cite{Herdeiro:2014goa,Herdeiro:2015gia}.\footnote{In the scalar case, $q$ is the fraction of the total angular momentum stored in the scalar field, $J^\Psi/J$. This follows from  $J^\Psi=mQ$. For the Proca field, there is a (numerically always small) correction to this interpretation, as the angular momentum stored in the field is not precisely quantized in terms of the Noether charge, in the presence of a horizon -- see Appendix C in~\cite{Herdeiro:2016tmi}.}  For each value of $M,J,m,n$, and fixed $\mu$, there can be (at most) three solutions: the Kerr solution, with $q=0$; the boson star solution, with $q=1$; and a HBH solution, with $0<q<1$. The normalized charge $q$ for the HBH solution is a function $M,J,m,n$ (and of course of the scalar/vector field mass $\mu$, which is a parameter of the theory).

\begin{figure*}[ht] \centering
\includegraphics[width=0.95\textwidth]{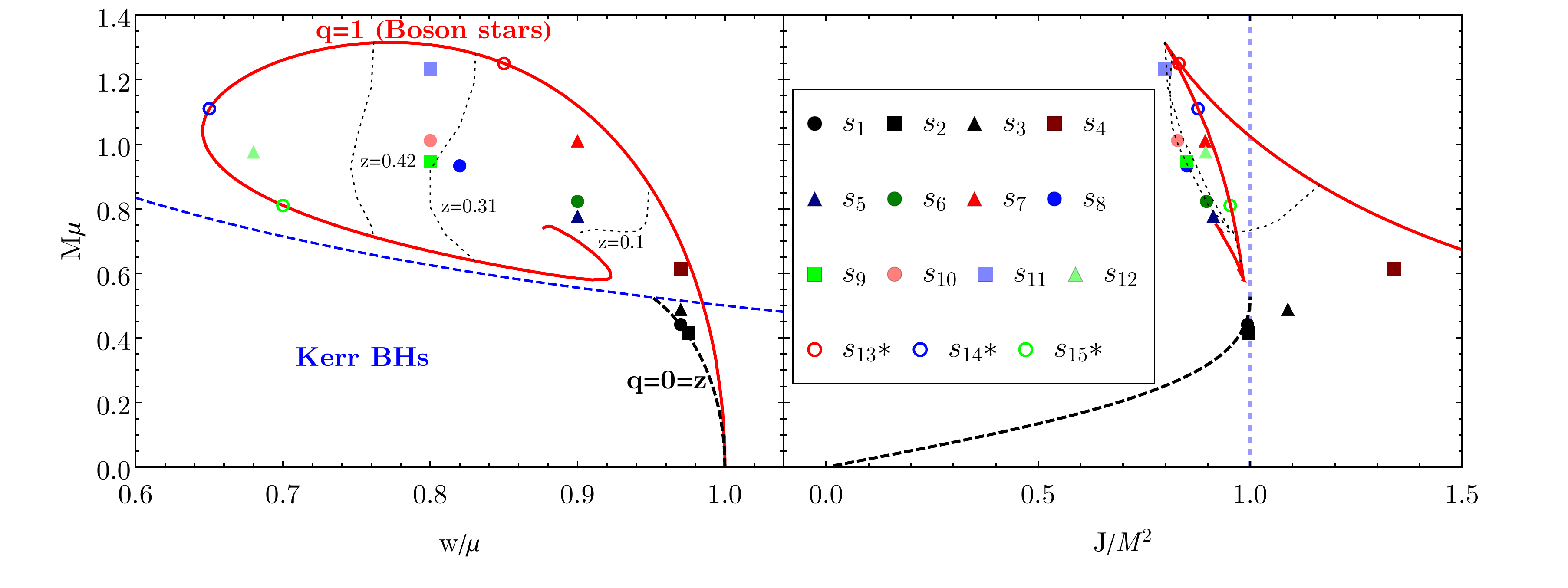}
\includegraphics[width=0.95\textwidth]{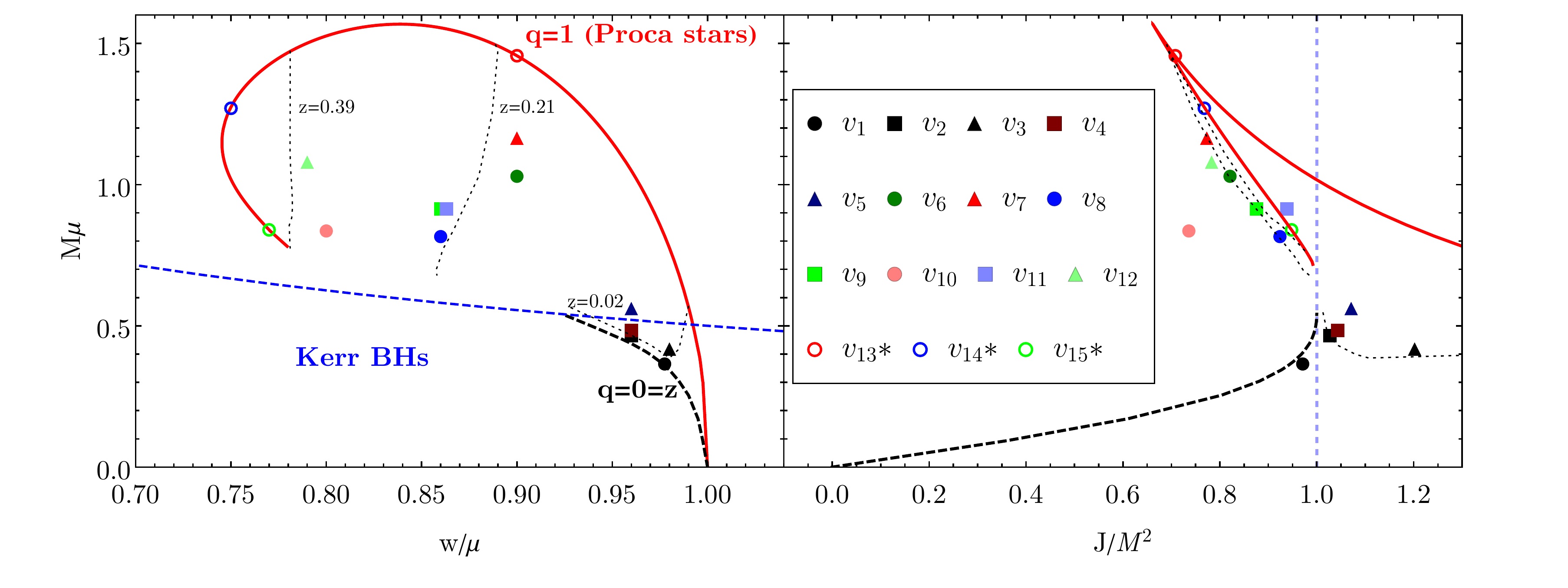}
\caption{Left panels: The the mass-frequency ($M\mu$-$w/\mu$) parameter space for spinning BHs with scalar (top panel) and vector (bottom panel) hair with azimuthal number $m=1$ (cf.~Ref.~\cite{Herdeiro:2015gia,Herdeiro:2016tmi}). The solutions considered in this work are denoted by markers. The properties of each solution are listed in Table~\ref{tab:parameters} below. The continuous red lines $(q=1)$ denote scalar/Proca boson stars. The dashed black curves ($q=z=0$) corresponds to Kerr BHs which are marginally stable against the superradiance instability. The thin dotted black curves denotes solutions of constant $z$ value [cf.~\eqref{zeta}].
Standard Kerr BHs exist below the dashed blue line, defined by the condition $a^\star<1$ or $M\mu<\frac{m}{2} \frac{\mu}{w}$.
Right panels: the corresponding spin-mass ($a^\star:=J/M^2$ -$M\mu$) parameter space. The markers and the curves are the same as in the left panels. In this case Kerr BHs exist only on the left of the dashed blue line, where $a^\star<1$.} \label{fig:parameterspace}
\end{figure*}

As discussed in Refs.~\cite{Herdeiro:2014goa,Herdeiro:2015gia}, stationary metric solutions only exist at the threshold of the superradiant condition~\cite{Superradiance}, i.e. when
\begin{equation}
w=m\Omega_H \,, \label{cloudcond}
\end{equation}
where $\Omega_H(M,J,q)$ is the angular velocity at the event horizon. Therefore, the frequency $w$ for the HBH solution is completely fixed in terms of $M$, $J$ and $q$ (cf.~Ref.~\cite{Herdeiro:2015gia} for details).

It is important to notice that the mass term $\mu$ enters the solutions simply as a rescaling factor. Hence, for fixed values of the spin parameter $a^\star$ and of the normalized Noether charge $q$, solutions with the same combination $M\mu$ are the same. In the following, we will present our results in terms of the dimensionless quantities
\begin{eqnarray}
 M\mu &\sim& 0.748\left(\frac{M}{10 M_\odot}\right) \left(\frac{m_b }{10^{-11}\, {\rm eV}}\right)\,, \label{masscale} \\
 \frac{\nu}{\mu} &\sim& 0.066\left(\frac{\nu}{{\rm kHz}}\right) \left(\frac{10^{-11}\, {\rm eV}}{m_b}\right)\,, \label{freqscale}
\end{eqnarray}
for the BH mass and for the geodesic frequencies, respectively. By using the above relations, it is straightforward to convert our results in physical units for a given boson mass $m_b=\mu \hbar$ expressed in eV.

The parameter space of HBH solutions is summarized in Fig.~\ref{fig:parameterspace}. The numerical solutions considered in this work (denoted by markers) are a representative sample of the entire parameter space. Some relevant properties of these solutions are listed in Table~\ref{tab:parameters}. In particular, note that some HBHs can exceed the Kerr bound on the spin, i.e.\ they have $a^\star>1$ without being naked singularities~\cite{Herdeiro:2014goa,Herdeiro:2015gia}.

\begin{table*}
\begin{footnotesize}
\centering
\begin{tabular}{C{0.3cm}C{0.3cm}C{0.3cm}C{0.3cm}C{0.3cm}C{0.3cm}C{0.3cm}C{0.3cm}C{0.3cm}C{0.7cm}C{1.1cm}C{0.7cm}}
\hline
\hline
Label & $z$ & $q$ & $M\mu$ & $a^\star$ & $M_B/M$ & $a_B^\star$ & $w/\mu$ & $r_H /M$ & $\rho_{\rm ISCO}/M$ & $\rho^{\rm{Kerr}}_{\rm ISCO}/M$ & $R_{99}/M$\\
\hline
\texttt{s1} & 0.0008 & 0.014 & 0.441 & 0.994 & 0.007 & 0.014 & 0.97 & 0.317 & 2.117 & 2.060 & 46.69 \\
\texttt{s2} & 0.0063 & 0.127 & 0.415 & 0.996 & 0.054 & 0.127 & 0.975 & 0.481 & 2.150 & 2.034 & 56.47 \\
\texttt{s3} & 0.0346 & 0.585 & 0.489 & 1.089 & 0.317 & 0.634 & 0.97 & 0.511 & 1.738 & ($a^\star>1$) & 46.40 \\
\texttt{s4} & 0.0584 & 0.988 & 0.615 & 1.339 & 0.818 & 1.334 & 0.97 & 0.325 & 0.850 & ($a^\star>1$) & 45.13 \\
\texttt{s5} & 0.1233 & 0.649 & 0.777 & 0.913 & 0.502 & 0.591 & 0.9 & 0.103 & 1.177 & 2.777 & 12.46 \\
\texttt{s6} & 0.1448 & 0.762 & 0.823 & 0.897 & 0.594 & 0.682 & 0.9 & 0.182 & 1.043 & 2.843 & 12.65 \\
\texttt{s7} & 0.1875 & 0.987 & 1.011 & 0.894 & 0.883 & 0.882 & 0.9 & 0.198 & 0.577 & 2.857 & 12.76 \\
\texttt{s8} & 0.2765 & 0.844 & 0.933 & 0.851 & 0.749 & 0.719 & 0.82 & 0.107 & 0.830 & 3.138 & 6.003 \\
\texttt{s9} & 0.3157 & 0.877 & 0.947 & 0.850 & 0.793 & 0.744 & 0.8 & 0.106 & 0.778 & 3.145 & 5.154 \\
\texttt{s10} & 0.3308 & 0.919 & 1.011 & 0.829 & 0.827 & 0.762 & 0.8 & 0.138 & 0.713 & 3.239 & 5.385 \\
\texttt{s11} & 0.3596 & 0.999 & 1.232 & 0.799 & 0.959 & 0.799 & 0.8 & 0.081 & 1.146 & 3.413 & 5.674 \\
\texttt{s12} & 0.5360 & 0.997 & 0.976 & 0.895 & 0.982 & 0.893 & 0.68 & 0.041 & 0.819 & 2.915 & 2.166 \\
\texttt{s13}$^*$ & 0.2775 & 1 & 1.250 & 0.834 & 1 & 0.834 & 0.85 & 0 & 2.473 & 3.277 & 8.029 \\
\texttt{s14}$^*$ & 0.5775 & 1 & 1.110 & 0.877 & 1 & 0.877 & 0.65 & 0 & 1.091 & 3.037 & 2.112 \\
\texttt{s15}$^*$ & 0.5100 & 1 & 0.811 & 0.953 & 1 & 0.953 & 0.7 & 0 & 0.212 & 2.532 & 1.389 \\
\hline
\texttt{v1} & 0.004 & 0.08 & 0.365 & 0.971 & 0.030 & 0.083 & 0.9775 & 0.679 & 2.367 & 2.201 & 72.73 \\
\texttt{v2} & 0.019 & 0.248 & 0.466 & 1.027 & 0.130 & 0.256 & 0.96 & 0.373 & 1.954 & ($a^\star>1$) & 40.38 \\
\texttt{v3} & 0.027 & 0.686 & 0.418 & 1.202 & 0.319 & 0.908 & 0.98 & 0.694 & 1.911 & ($a^\star>1$) & 76.61 \\
\texttt{v4} & 0.028 & 0.370 & 0.485 & 1.043 & 0.203 & 0.388 & 0.96 & 0.387 & 1.844 & ($a^\star>1$) & 39.85 \\
\texttt{v5} & 0.053 & 0.677 & 0.561 & 1.071 & 0.426 & 0.725 & 0.96 & 0.417 & 1.497 & ($a^\star>1$) & 38.08 \\
\texttt{v6} & 0.180 & 0.949 & 1.029 & 0.821 & 0.825 & 0.774 & 0.9 & 0.185 & 0.674 & 3.264 & 13.75 \\
\texttt{v7} & 0.188 & 0.988 & 1.164 & 0.773 & 0.888 & 0.758 & 0.9 & 0.163 & 0.521 & 3.503 & 13.34 \\
\texttt{v8} & 0.229 & 0.88 & 0.816 & 0.924 & 0.802 & 0.804 & 0.86 & 0.067 & 0.872 & 2.717 & 10.08 \\
\texttt{v9} & 0.232 & 0.91 & 0.915 & 0.875 & 0.821 & 0.791 & 0.863 & 0.098 & 0.764 & 3.006 & 10.23 \\
\texttt{v10} & 0.260 & 0.999 & 1.254 & 0.736 & 0.922 & 0.727 & 0.86 & 0.116 & 0.434 & 3.689 & 9.412 \\
\texttt{v11} & 0.360 & 0.999 & 0.836 & 0.938 & 0.981 & 0.917 & 0.8 & 0.024 & 0.393 & 2.635 & 6.639 \\
\texttt{v12} & 0.372 & 0.99 & 1.173 & 0.783 & 0.970 & 0.780 & 0.79 & 0.051 & 0.316 & 3.507 & 6.198 \\
\texttt{v13}$^*$ & 0.190 & 1 & 1.454 & 0.708 & 1 & 0.708 & 0.9 & 0 &  none & 3.861 & 12.75 \\
\texttt{v14}$^*$ & 0.438 & 1 & 1.274 & 0.768 & 1 & 0.768 & 0.75 & 0 & none & 3.600 & 5.982 \\
\texttt{v15}$^*$ & 0.407 & 1 & 0.840 & 0.948 & 1 & 0.948 & 0.77 & 0 & none & 2.571 & 7.060 \\
\hline
\hline
\end{tabular}
\caption{Properties of the HBH solutions and of the boson/Proca stars considered in this work. Scalar and vector solutions are denoted by \texttt{sX} and \texttt{vX} respectively ($X=1,2,3,...$). Labels with a star denote boson/Proca stars.
HBH solutions are ordered by growing values of $z=q(1-w^2/\mu^2)$ [see discussion around Eq.~\eqref{zeta}].
$M_B$ and $a_B^\star:= J_B/M^2$ denote the mass and the (dimensionless) angular momentum of the bosonic condensate as computed in Refs.~\cite{Herdeiro:2014goa,Herdeiro:2015waa}, whereas $r_H$ and $R_{99}$ are the BH horizon and the radius containing 99\% of the bosonic mass~\cite{Liebling:2012fv}. $\rho_{\rm ISCO}$ is the circumferential radius [$\rho:= re^{F_2(r)}$, cf. line element~\ref{metric}] of the ISCO, and $\rho^{\rm{Kerr}}_{\rm ISCO}$ is the corresponding value for a Kerr BH with the same mass and spin (provided $a^\star<1$).
\label{tab:parameters}}
\end{footnotesize}
\end{table*}

The parameter space of HBHs is very rich and there is no single known parameter that captures all the deviations from the Kerr case~\cite{Herdeiro:2015gia}. The normalized Noether charge $q$, which vanishes for Kerr, has certainly some information about these deviations; but alone it is not always a faithful indicator of the departures from Kerr, since experience shows the latter do not grow monotonically with $q$. To illustrate the limitations of $q$, consider a solution with a fairly large $q$ but with a very spread out bosonic hair, with respect to the horizon scale. In this case, the energy density of the bosonic field (and, in turn, its backreaction on the metric) is small. Thus, it is expected that such models can be very similar to Kerr in terms of the near horizon physics, despite the large $q$. To propose an improved deviation parameter, we observe that a measure of how much the bosonic field is spread out is provided by its asymptotic decay, which reads, say for the scalar field, $\Psi\sim e^{-\sqrt{\mu^2-w^2}r}/r$. The factor $\lambda\sim 1/\sqrt{\mu^2-w^2}$ is the asymptotic decay scale of the bosonic field, and hence provides a measure of  how much the latter is diluted. With this motivation, we found it convenient to introduce another \emph{phenomenological} parameter,
\begin{equation}
 z:=q(1-w^2/\mu^2)\,,\label{zeta}
\end{equation}
in terms of which the deviations of the BH solutions indeed display a smoother behavior, cf.\ Table~\ref{tab:parameters} and the discussion in the next sections. In particular, in the analyses presented below, we found that for larger $z$, it is easier to distinguish HBHs (both in the scalar and in the vector case) from their Kerr counterpart. From the definition~\eqref{zeta}, we note that $z$ is small when $q\sim0$
(i.e. close to vacuum Kerr BHs),
but also when $w\sim\mu$, i.e. in the rightmost lower part of the parameter space shown in the left panels of Fig.~\ref{fig:parameterspace}, in which $q$ can also be close to unity but the bosonic field is non-compact, and the boson/Proca stars are in the Newtonian regime.

\section{Quasi-periodic oscillations in the X-ray flux of accreting
 black holes}\label{sec:QPO}
QPOs in the X-ray spectrum from accreting BHs and neutron stars provide a clean environment to study the
properties of strong gravitational fields through the motion of matter that generates these oscillations (for some reviews see \cite{vanderklis:2000,Motta:2016vwf}).
In the last decades, a few models have been developed to describe the QPO spectra. Among them, especially successful models are the RPM and the epicyclic resonance model~\cite{Kluzniak_Abramowicz:2001,Abramowicz_Kluzniak:2001}. In this work we base our analysis upon the RPM, although qualitatively similar results apply to any QPO model based on geodesic motion~\cite{Maselli:2017kic}.


The RPM was originally formulated to model the QPO triplets, formed by twin QPOs around $\sim1$~kHz and a low-frequency
QPO mode (the so-called Horizontal Branch Oscillations), observed in the X-ray spectra from accreting neutron
stars~\cite{Stella_Vietri:1998,Stella:1998mq}. The high-frequency twin QPOs are identified with the \textit{azimuthal frequency}
($\nu_\varphi$) and the \textit{periastron precession frequency} ($\nu_\tn{per} = \nu_\varphi-\nu_r$) of matter in a
quasi-circular orbits with a given radius $r$, while the low-frequency QPO mode is identified with the \textit{nodal precession
frequency} ($\nu_\tn{nod}=\nu_\varphi - \nu_\theta$). Here $\nu_r$, $\nu_\theta$ are the radial and vertical epicyclic
frequencies of the quasi-circular orbit, respectively. We remark that the three QPOs at frequencies $\nu_\varphi$,
$\nu_\tn{per}$ and $\nu_\tn{nod}$ are emitted at the same radius $\bar{r}$. Different QPO triplets observed from individual
neutron stars are well reproduced in the RPM, since they are assumed to correspond to different emission radii.

The RPM has been applied to BH systems~\cite{Stella_V_M:1999}, but only recently, when a QPO triplet has been detected in the accreting BH system, a complete application of the RPM model to BHs has been possible. Two high-frequency and one low-frequency QPOs have been simultaneously detected from GRO J1655-40~\cite{2005ApJ...629..403C,Motta:2013wga}. Their centroid frequencies have been measured by the Rossi X-ray Timing Explorer (RXTE) mission with 1$\sigma$ uncertainties in the 1-2\% range:
\begin{equation}
\label{freq}
\nu_\varphi= 441^{+2}_{-2}~\tn{Hz},\
\nu_\tn{per}= 298^{+4}_{-4}~\tn{Hz},\
\nu_\tn{nod}= 17.3^{+0.1}_{-0.1}~\tn{Hz}~.
\end{equation}
By fitting this QPO triplet with the RPM frequencies corresponding to the Kerr metric, it has been possible to obtain
precise values of the BH mass and spin, $M=(5.31\pm 0.07) M_\odot$ and $a^\star = 0.290 \pm 0.003$, and the value of the
radius from which the QPO triplet originated. The value of the mass is fully consistent with the value inferred from
optical/NIR spectro-photometric observations, $(5.4\, \pm 0.3) M_\odot$ \cite{Beer_Podsiadlowski_2002}. The QPO radius has been found to be $\bar{r}=(5.68 \,\pm 0.04)\, M$, very deep in the BH gravitational field.

From the detection of a single QPO triplet it is only possible to extract the three quantities $M,a^\star,\bar{r}$, with no redundancy.  Detections of more triplets from the same BH can give additional information on the properties of strong-field gravity, such as, for instance, the radial dependence of the azimuthal and epicyclic frequencies.  Since
the signal-to-noise ratio of incoherent signals like QPOs scales linearly with the rates of photons detected by counting instruments, the precision in QPO measurements increases with the area of the detector. Large-area detectors, such as the future eXTP satellite~\cite{eXTP,Zhang:2016ach} and the proposed LOFT detector~\cite{feroci2012large}, will allow the detection of QPOs from accreting BHs with much larger accuracy. In this article, our calculations are based on the large area instrument LAD/eXTP, which is expected to be
$\sim 5$ times more accurate than RXTE-PCA. Thus, for instance, we expect that a measurement with eXTP would have errors $\sim5$ times smaller than the RXTE errors reported in Eq.~\eqref{freq}.

\subsection{Azimuthal and epicyclic frequencies of a spinning compact object}\label{sec:epyc}
If the extra bosonic field is only weakly coupled to ordinary matter, it does not interfere with the dynamics of test bodies, except for its own gravitational potential. Thus, the motion of test particles on the equatorial plane is geodesic and can be conveniently studied in terms of the epicyclic frequencies ($e.g.$\
\cite{wald2010general,1999MNRAS.304..155M,Abramowicz:2002iu,Vincent:2013uea}). Here we follow the derivation of~\cite{Maselli:2014fca} [cf.\ also Refs.~\cite{2005Ap&SS.300..127A} for earlier work] for the epicyclic frequencies of a stationary and axisymmetric geometry such as~\eqref{metric}. Namely, we consider
\begin{equation}
ds^2=g_{tt}dt^2+g_{rr}dr^2+g_{\theta\theta}d\theta^2+2g_{t\varphi}dtd\varphi +g_{\varphi\varphi}d\varphi^2\,,\label{genmetric}
\end{equation}
where $g_{\mu\nu}=g_{\mu\nu}(r,\theta)$, the angular velocity of corotating orbits reads
\begin{equation}
 \Omega:=2\pi \nu_\varphi =\frac{-\partial_r g_{t\varphi}+\sqrt{(\partial_r g_{t\varphi})^2-\partial_r g_{tt}\partial_r g_{\varphi\varphi}}}{\partial_r g_{tt}}\,, \label{Omega}
\end{equation}
whereas the radial and angular frequencies are
\begin{eqnarray}
\nu_r^2&=&\frac{(g_{tt}+\Omega
g_{t\varphi})^2}{2(2\pi)^2g_{rr}}\frac{\partial^2{\cal U}}{\partial
r^2}\left(r,\frac{\pi}{2}\right)\,,\nonumber\\
\nu_\theta^2&=&\frac{(g_{tt}+\Omega
g_{t\varphi})^2}{2(2\pi)^2g_{\theta\theta}}\frac{\partial^2{\cal
U}}{\partial\theta^2} \left(r,\frac{\pi}{2}\right)\,.\label{epic}
\end{eqnarray}
Here we have defined
\begin{equation}
{\cal U}(r,\theta)=g^{tt}-2lg^{t\varphi}+l^2g^{\varphi\varphi}\,,\label{defU}
\end{equation}
with $l:=L/E$, and
\begin{eqnarray}
 E&=& -\frac{g_{tt}+g_{t\varphi}\Omega}{\sqrt{g_{tt}-2g_{t\varphi}\Omega-g_{\varphi\varphi}\Omega^2}}\,,\\
 L&=& \frac{-g_{t\varphi}+g_{\varphi\varphi}\Omega}{\sqrt{g_{tt}-2g_{t\varphi}\Omega-g_{\varphi\varphi}\Omega^2}} \,,
\end{eqnarray}
are the specific energy and angular momentum of the corotating orbit. Note that ${\cal U}$ is related to the standard effective potential $V(r)$ for radial, equatorial motion,
\begin{equation}
 {\dot r}^2=V(r):=-g_{rr}^{-1}\left[1+E^2{\cal U}(r,\pi/2)\right]\,,
\end{equation}
where a dot denotes derivative with respect to the proper time.
In particular, circular orbits at $r=\bar{r}$ require $E^2=-1/{\cal U}(\bar{r},\pi/2)$ (and therefore ${\cal U}(\bar{r},\pi/2)<0$) as well as $\frac{\partial {\cal U}}{\partial r}(\bar{r},\pi/2)=0$.
The ISCO is defined by the further condition $\frac{\partial^2 {\cal U}}{\partial r^2}(r_{\rm ISCO},\pi/2)=0$ (i.e., $\nu_r=0$), and therefore the stability of circular orbits requires $\frac{\partial^2 {\cal U}}{\partial r^2}(\bar{r},\pi/2)>0$ (i.e., $\nu_r>0$).

Finally, the periastron
and nodal precession frequencies read
\begin{equation}
\nu_{\tn{nod}}=\nu_{\varphi}-\nu_{\theta}\quad \ ,\quad
\nu_{\tn{per}}=\nu_{\varphi}-\nu_{r}\,.
\end{equation}
In several models, the QPO frequencies are related to the epicyclic frequencies defined above. In particular, in the RPM, the azimuthal frequency, $\nu_\varphi$, defined in Eq.~\eqref{Omega}, and the two above frequencies are identified with a QPO triplet.

\section{Results}\label{sec:results}

\begin{figure*}[ht] \centering
\includegraphics[width=1.05\textwidth]{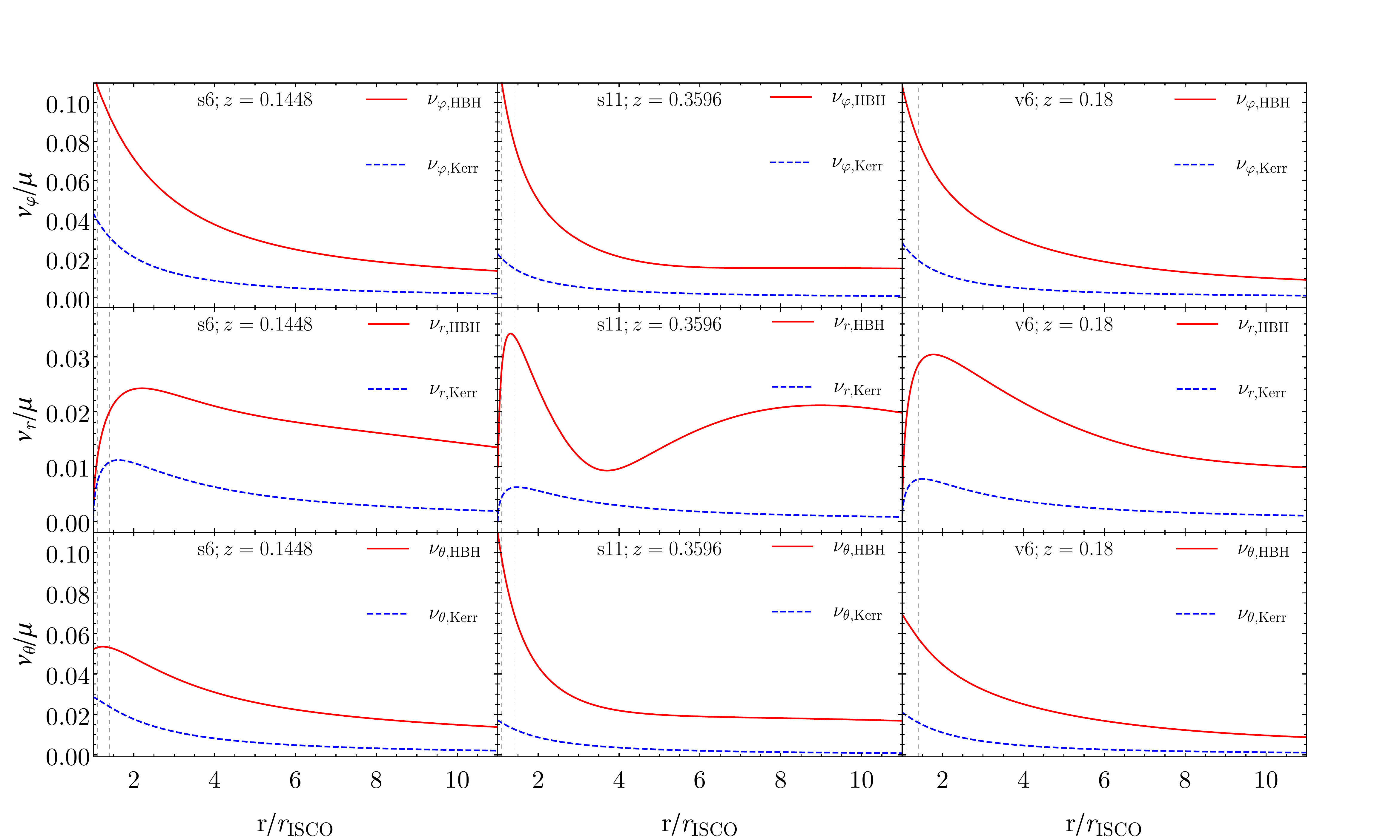}
\caption{Azimuthal and epicyclic frequencies 
$(\nu_\varphi,\nu_r,\nu_\theta)$ of a spinning BH with scalar hair (first two columns, corresponding to solutions \texttt{s6} and \texttt{s11}) and vector hair (last column, corresponding to solution \texttt{v6}), compared with the corresponding frequencies for a Kerr BH with the same mass $M$ and spin $J$. The radial coordinate is normalized by the ISCO of the corresponding solution. The vertical dashed lines denote the radii $\bar{r}=1.1\,r_{\rm ISCO}$ and $\bar{r}=1.4\,r_{\rm ISCO}$ that we have considered in the $\chi^2$ analysis of Fig.~\ref{fig:chi2}.
\label{fig:frequencies}}
\end{figure*}
\subsection{Azimuthal and epicyclic frequencies of BHs with bosonic hair}

We have computed the frequencies $\nu_\varphi$, $\nu_r$,
$\nu_\theta$, $\nu_{\tn{per}}$ and $\nu_{\tn{nod}}$ for the HBH solutions presented in Fig.~\ref{fig:parameterspace} and in Table~\ref{tab:parameters}.
In Fig.~\ref{fig:frequencies}, we show the azimuthal and epicyclic frequencies for some representative example of solutions, and we compare them with their Kerr counterparts, which are given by~\cite{2001ApJ...548..335S,PhysRevD.5.814,1987PASJ...39..457O}
\begin{eqnarray}
\nu_\varphi^{\tn{Kerr}}&=&\frac{1}{2\pi}\frac{M^{1/2}}{R^{3/2}+ {a^\star}
M^{3/2}}\,, \label{nu1GR}\\
\nu_r^{\tn{Kerr}}&=&\nu_\varphi^{\tn{Kerr}}\sqrt{1-\frac{6M}{R}+
8{a^\star}\frac{ M^{3/2}}{R^{3/2}}- 3{a^\star}^2\frac{
M^2}{R^2}}\,,\label{nu2GR}\\
\nu_\theta^{\tn{Kerr}}&=&\nu_\varphi^{\tn{Kerr}}\sqrt{1-4{a^\star}\frac{
M^{3/2}}{R^{3/2}}+ 3{a^\star}^2\frac{ M^2}{R^2}}\,,
\label{nu3GR} \end{eqnarray}
where $R$ is the standard Boyer-Lindquist radial coordinate, which is related to the radial coordinate defined in Eq.~\eqref{metric} by a simple shift, $r=R-{a^\star}^2 M/(1+\sqrt{1-{a^\star}^2})$~\cite{Herdeiro:2015gia}.
When $a^\star<1$, we have performed this transformation in order to evaluate both metrics in the same coordinate system.

In Figure~\ref{fig:frequencies}, we show that the behavior of the azimuthal and epicyclic frequencies of the HBH solutions can be drastically different from that of a Kerr BH with the same mass and spin. In particular, the geodesic frequencies near the ISCO display order-one deviations and, as shown in Table~\ref{tab:parameters}, the circumferential radius of the ISCO itself can differ by more than a factor of $5$~\cite{Herdeiro:2014goa,Herdeiro:2015gia}.

\begin{figure}[ht] \centering
\includegraphics[width=0.48\textwidth]{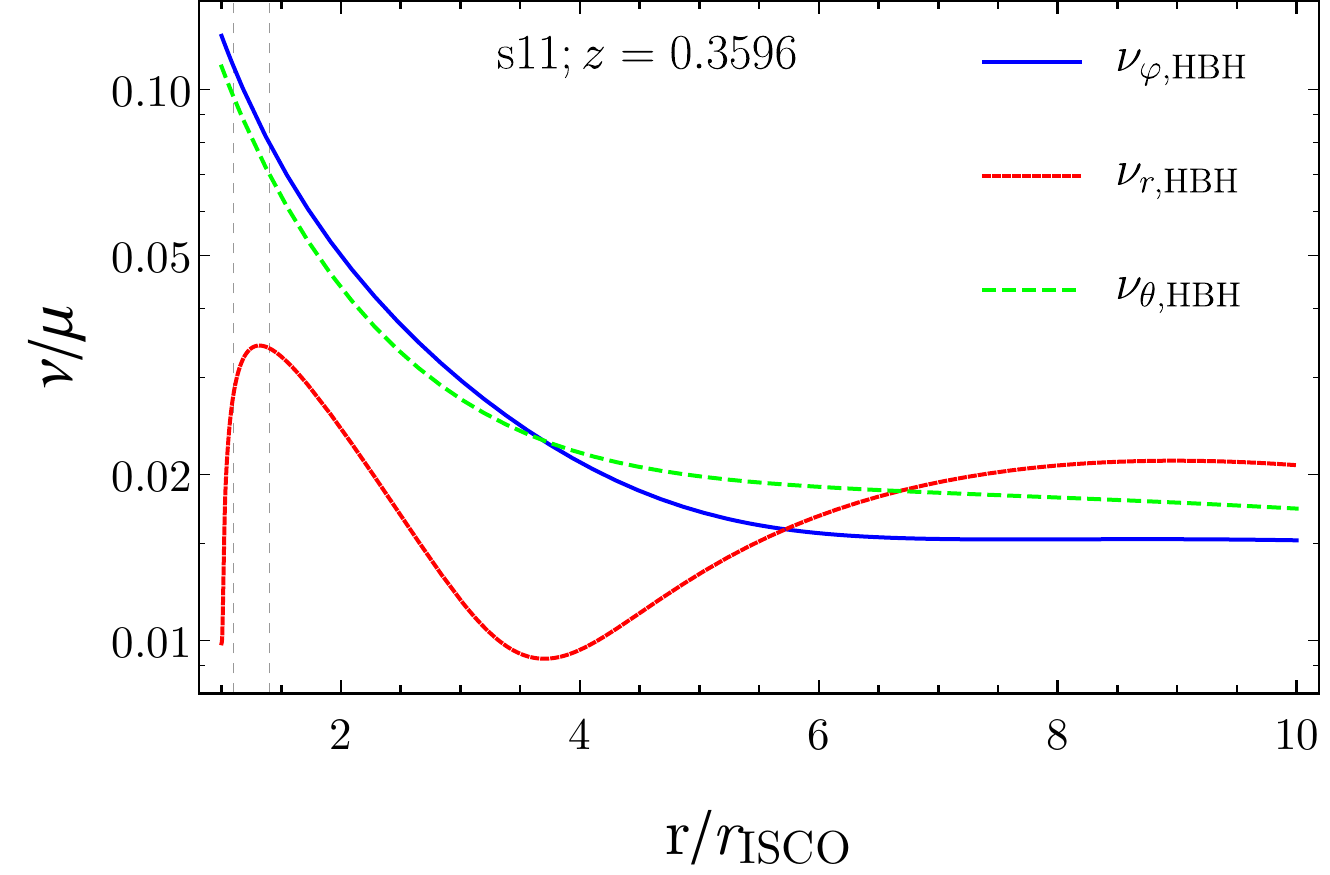}
\caption{Azimuthal and epicyclic frequencies 
$(\nu_\varphi,\nu_r,\nu_\theta)$ of the spinning BH with scalar hair solution \texttt{s11}. Note that the hierarchy of the frequencies depends on the emission radius.
\label{fig:allfreq}}
\end{figure}

For clarity, in Figure~\ref{fig:allfreq} we show the azimuthal and epicyclic frequencies for solution \texttt{s11}. It is interesting to note that the hierarchy of the frequencies depends on the emission radius.

We stress that in the theory under consideration [cf.\ Eq.~\eqref{theory}] the boson-mass term $\mu$ is a scale factor for all dimensionful quantities. In particular, we found it convenient to normalize the frequencies shown in Figs.~\ref{fig:frequencies} and \ref{fig:allfreq} by $\mu$. By using Eq.~\eqref{freqscale}, it is straightforward to convert the values of $\nu/\mu$ to the corresponding frequency expressed in Hertz, after a given mass $\mu$ is specificed. With our normalization, the results presented in Figs.~\ref{fig:frequencies}, \ref{fig:allfreq} and below are valid for any value of $\mu$.

\subsection{QPO spectroscopy of BHs with bosonic hair}\label{sec:analysisBH}

According to the RPM, the three simultaneous QPO frequencies
$(\nu_\varphi,\nu_{\tn{per}},\nu_{\tn{nod}})$ are
all generated at the same radial coordinate $\bar{r}$ in the accretion flow.
In the Kerr case, for each such triplet, Eqs.~\eqref{nu1GR}--\eqref{nu3GR} can be solved analytically for the three unknown parameters $(M,{a^\star},\bar r)$.
However, as discussed above, HBHs have an extra parameter. Therefore, at least
one more QPO triplet is required in the RPM in order to develop a null-hypothesis test of the Kerr metric.
Furthermore, as previously explained solutions with the same ($M\mu$, $a^\star$, $q$) are equivalent, so it is sufficient to set the value of $\mu$ to make the mass $M$ agree with the observations.

It is expected that future very high
effective-area instruments will allow for the detection of
multiple QPO triplets corresponding to different
radii in individual stellar mass BHs, which can therefore provide several independent tests. In this section, following the framework developed in Ref.~\cite{Maselli:2014fca}, we explore the potential of such
observations to discriminate HBHs against Kerr BHs in the strong-field gravitational regime.

\begin{figure}[h!]
\centering
\includegraphics[width=0.48\textwidth]{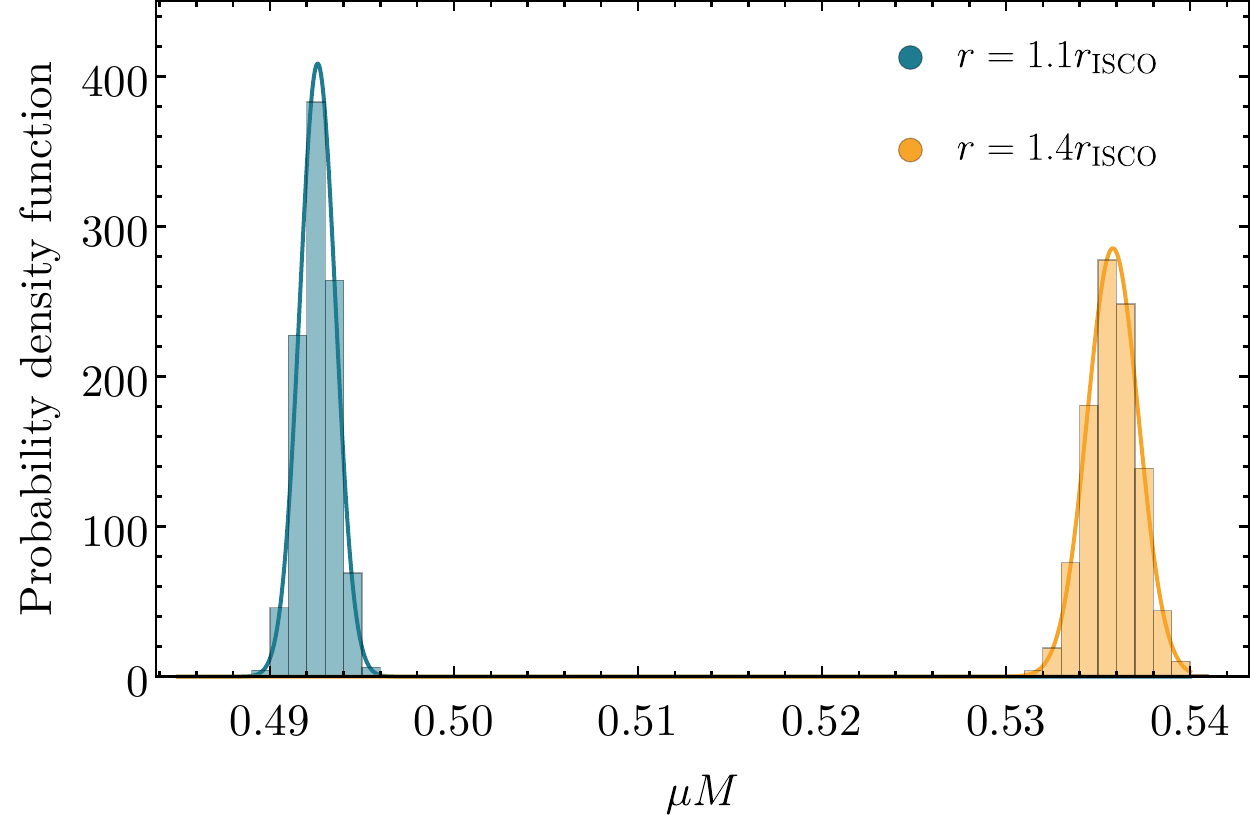}
\includegraphics[width=0.49\textwidth]{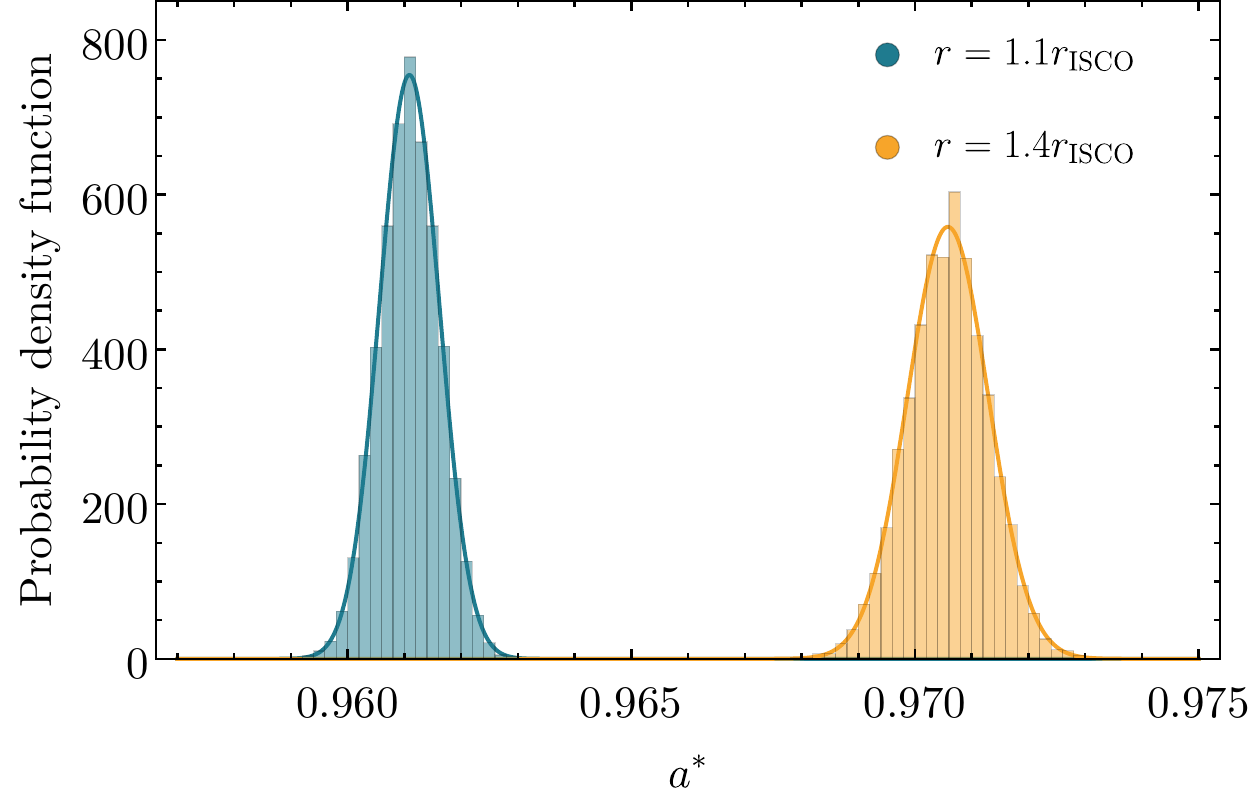}
\caption{Distribution of the BH mass $M$ (top panel) and spin $a^\star$ (bottom panel) as obtained from the measurement of the azimuthal and epicyclic frequencies $(\nu_\varphi,\nu_{\tn{per}},\nu_{\tn{nod}})$ at two different radii for the solution \texttt{s10} in Table~\ref{tab:parameters}. The generated frequencies follow Gaussian distributions with
relative widths 5 times lower than those measured with the RXTE/PCA
from GRO~J1655-40, as expected for the eXTP/LAD instrument.
}
\label{fig:distribution}
\end{figure}

We concentrate first on the case in which two different QPO
triplets are measured. We proceed as follows:

\begin{enumerate}
\item
We assume the source of the QPOs is an accreting BH with bosonic hair, and choose a specific solution of Table~\ref{tab:parameters}. From the tabulated solution, we can extract the mass, spin, Noether charge, and all relevant parameters, as explained in Refs.~\cite{Herdeiro:2015gia,Herdeiro:2016tmi}.
\item For a given solution, we compute numerically two triplets
$\nu_{\tn{ref1}}=(\nu_\varphi,\nu_{\tn{per}},\nu_{\tn{nod}})_1$ and
$\nu_{\tn{ref2}}=(\nu_\varphi,\nu_{\tn{per}},\nu_{\tn{nod}})_2$,
corresponding to two different fiducial emission radii,
$r_1/r_{\tn{ISCO}}=1.1$ and $r_2/r_{\tn{ISCO}}=1.4$, respectively\footnote{We have checked that changing the emission radii in the near-ISCO region does not affect our results significantly.}.
As previously discussed, we
assume that these are the QPO frequencies measured by a next-generation large-area instrument (e.g.\ eXTP), with corresponding relative uncertainties $5$ times smaller than those measured with
the RXTE-PCA for GRO~J1655-40 [see Eq.~\eqref{freq}].
In practice, we consider that in the relevant range the QPO frequencies will be detected with $\approx 0.2\%$ relative accuracy. We denote the corresponding absolute errors on the frequencies as $(\sigma_\varphi,\sigma_{\tn{per}},\sigma_{\tn{nod}})$.

\item
We then interpret these simulated data as if they were generated from the accretion disk of a standard Kerr BH,
and solve Eqs.~\eqref{nu1GR}--\eqref{nu3GR} to infer
the values of $(M_j,a^\star_j,\bar r_j)$, with $j=1,2$ corresponding
to the two QPO triplets.
\end{enumerate}

If the triplets
$\nu_{\tn{ref1}}, \nu_{\tn{ref2}}$ were
generated by a Kerr BH (i.e. with $q=0=z$), this procedure would yield
the same values of the mass and spin parameters ($M_1=M_2$ and
$a^\star_1=a^\star_2$), to within statistical and numerical
uncertainties.  Conversely, when $q\neq0$, we expect that this procedure would yield different values $M_1\neq M_2$ and $a^\star_1\neq a^\star_2$.

To quantify this discrepancy, we use the same Monte Carlo approach developed in Ref.~\cite{Maselli:2014fca}. For each solution in Table~\ref{tab:parameters}, we compute numerically $\nu_{\tn{ref1}}$ and $\nu_{\tn{ref2}}$ and then draw $N=10^4$ values
$(\nu_\varphi,\nu_{\tn{per}},\nu_{\tn{nod}})_{j}$ (with $j=1,2$ for $\nu_{\tn{ref1}}$ and $\nu_{\tn{ref2}}$, respectively),
from a Gaussian distribution, whose mean value is $\nu_{\tn{ref1}}$ or $\nu_{\tn{ref2}}$ and with standard deviation $(\sigma_\varphi,\sigma_{\tn{per}},\sigma_{\tn{nod}})$.
Then, by inverting Eqs.~\eqref{nu1GR}--\eqref{nu3GR}, we compute the corresponding $2N$ values of
$(M,a^\star,\bar r)$, which would correspond to the mass, spin and emission radius if the source was a Kerr BH.
It is straightforward to check that, if $N$ is sufficiently large, these quantities follow two multivariate Gaussian
distributions $\mathcal{N}_{1}({M}_1,{a}^\star_1,{r}_1,{\Sigma}_{1})$ and
$\mathcal{N}_{2}({M}_2,{a}^\star_2,{r}_2,{\Sigma}_{2})$,
where $(M_j,a^\star_j,r_j)$
are the expectation values of the two sets of parameters ($j=1,2$), whereas
${\Sigma}_j$ are their covariance matrices.

In Figure~\ref{fig:distribution}, we
show the distribution of mass (top panel) and spin (bottom panel)
obtained by this procedure, for solution \texttt{s10} in Table~\ref{tab:parameters}.
For this specific solution, it is already clear from Fig.~\ref{fig:distribution} that the distributions are grossly incompatible with the null hypothesis $M_1=M_2$, $a^\star_1=a^\star_2$. This analysis suggests that a measurement of two QPO triplets with the typical accuracy of a next-generation large effective-area detector can be used to distinguish HBHs from their counterparts in a very clear way.
In order to quantify this statement, we perform a $\chi^2$-analysis similar to that developed in \cite{Maselli:2014fca}. Namely, we construct the $\chi^2$ distributed variable with 3 degrees of freedom,
\begin{equation}
\chi^{2}=(\vec{x}-\vec{p})^{\tn{T}}{\Sigma}^{-1}(\vec{x}-\vec{p});
\end{equation}
where $\vec{p}=(\Delta M, \Delta a^{\star},\Delta \bar r)$, $\Delta M=M_1-M_2$, $\Delta a^{\star}=a^\star_1-a^\star_2$, and $\Delta \bar  r=\bar  r_1-\bar r_2$.
%
%
The condition $\chi^2=3.53,8.03,14.16$ defines the regions of $\Delta M$, $\Delta a^{\star}$, $\Delta \bar r$ that correspond to  $1\sigma$, $2\sigma$, and $3\sigma$ confidence level in a Gaussian distribution equivalent, respectively.
%

\begin{figure*}[ht]
\centering
\includegraphics[width=0.95\textwidth]{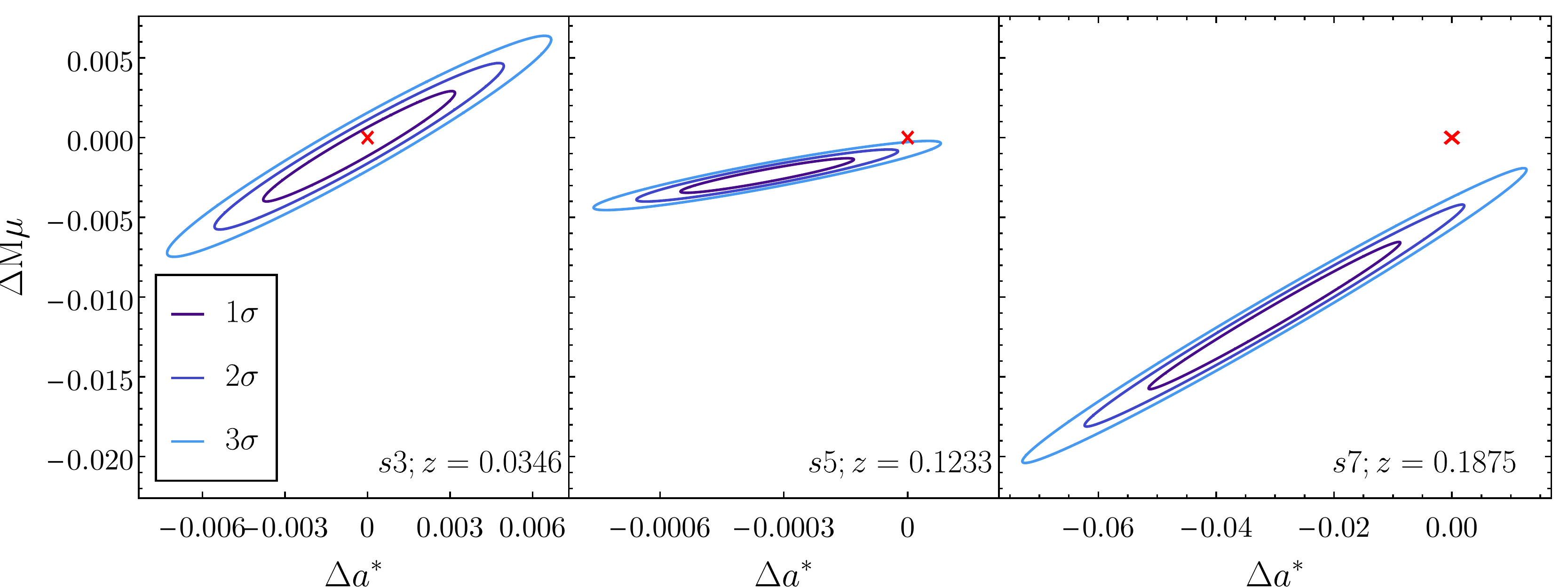}
\includegraphics[width=0.95\textwidth]{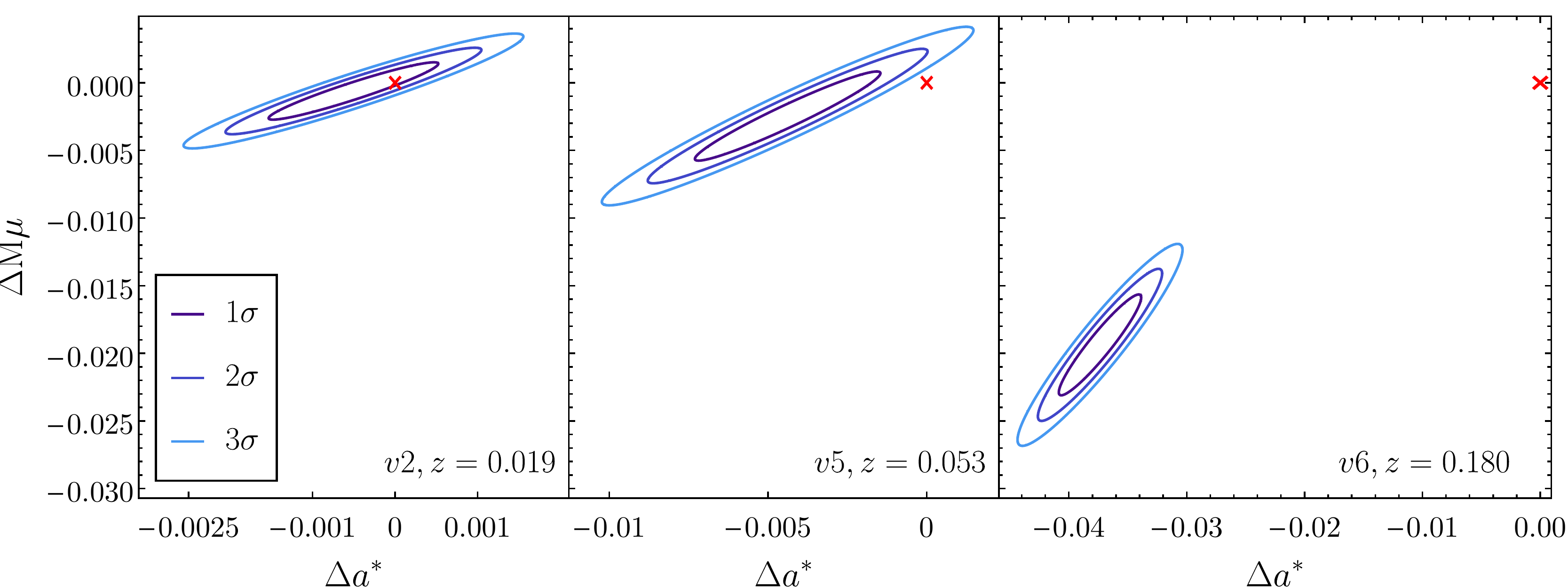}
\caption{The confidence levels  with which Kerr BHs can be tested against BHs with scalar (top row) and Proca hair (bottom row) hair are plotted in the $(\Delta M,\Delta a^\star)$ plane (see text for details).
Each panel shows a different solution listed in Table~\ref{tab:parameters}.
The red cross is the origin of the plane and corresponds to the Kerr case.
}
\label{fig:chi2}
\end{figure*}

In Figure~\ref{fig:chi2},
we show the results of this statistical analysis applied to the case
when the two simulated QPO triplets $(\nu_\varphi,\nu_{\tn{per}},\nu_{\tn{nod}})_{1,2}$, are computed assuming some representative solution of Table~\ref{tab:parameters}.
Each panel shows --~for a different solution~-- the regions in the parameter space $(\mu\Delta M,\Delta a^*)$ which correspond to $1\sigma$, $2\sigma$, and $3\sigma$
confidence level (top and bottom rows correspond to the scalar and the vector case, respectively).  The red cross denotes $\Delta M=0=\Delta a^*$, i.e. it identifies the point of the parameter space in which the solution would be compatible to a Kerr BH.
The two left panels correspond to solutions which are close to their Kerr counterpart (cf. Fig.~\ref{fig:parameterspace} and Table~\ref{tab:parameters}) and, accordingly, the red cross falls within the $1\sigma$ ellipse. The two middle panels correspond to the case in which the red cross is marginally outside the $3\sigma$ confidence level.
As anticipated, the deviations grow monotonically for increasing values of $z$ (although not shown, we have checked this statement for all solutions listed in Table~\ref{tab:parameters}). In particular, solutions with $z$ larger than that considered in the two middle panels of Fig.~\ref{fig:chi2} are all well outside the $3\sigma$ confidence level, as shown in the right panel for two representative cases.
This analysis indicates that --~if the RPM provides an accurate description of the QPO phenomenology~-- BHs with scalar hair with $z\gtrsim0.12$ and with vector hair with $z\gtrsim 0.025$ could be excluded/detected by future instruments at more than $3\sigma$ level.

\begin{figure*}[ht]
\centering
\includegraphics[width=0.8\textwidth]{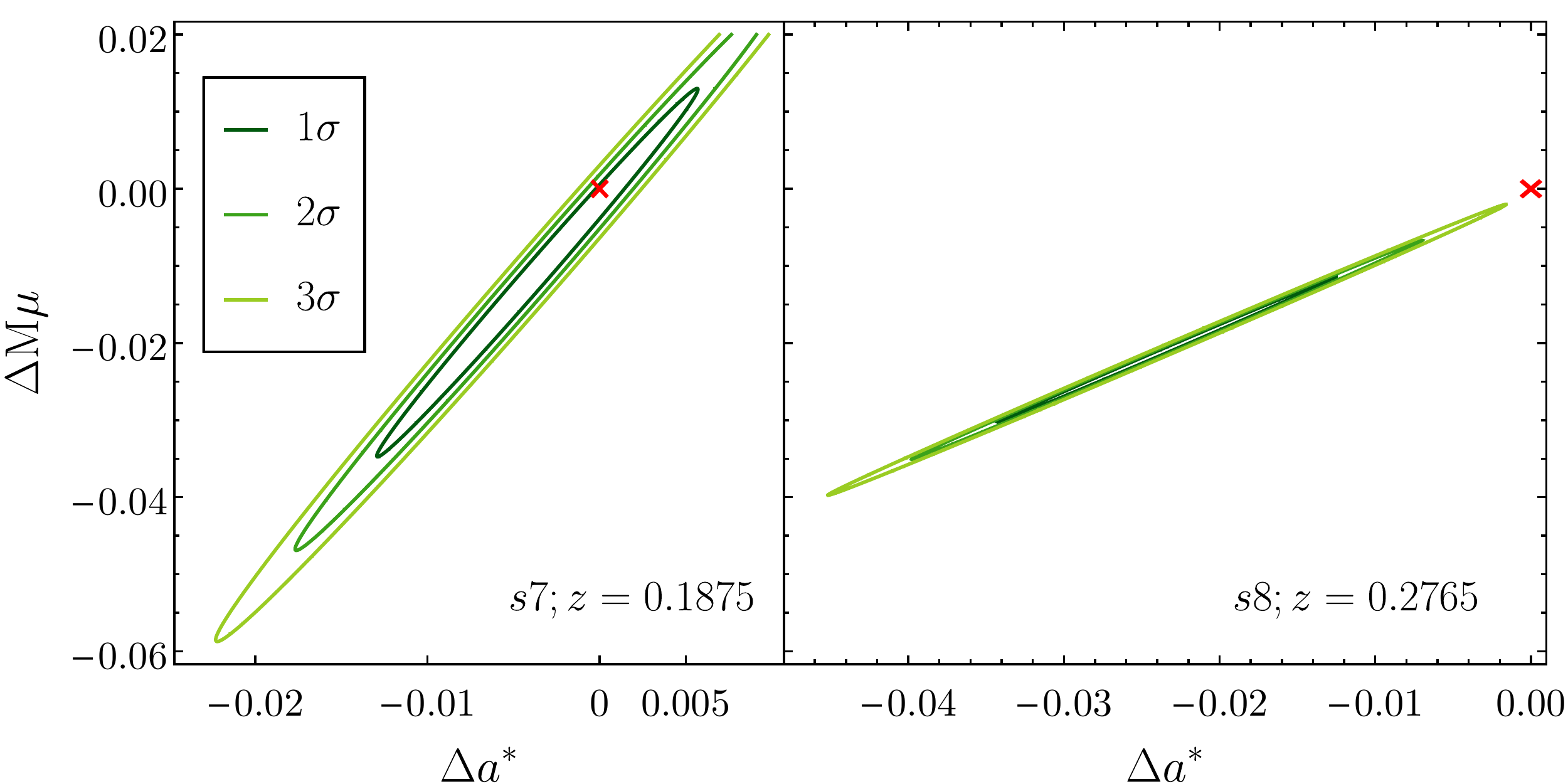}
\caption{The same as in Fig.~\ref{fig:chi2} but for a relative error in the measured QPO frequencies at the percent level, typical of RXTE. The left and right panels show solution \texttt{s7} and \texttt{s8}, respectively.}
\label{fig:chi2_RXTE}
\end{figure*}

It is also interesting to investigate whether current facilities can already put constraints on these models. By repeating the above analysis for the typical uncertainties of RXTE (at the level of $1\%$), we obtain two representative examples shown in Fig.~\ref{fig:chi2_RXTE}. These results show that BHs with scalar hair with $z\gtrsim 0.27$ can be constrained at least at $3\sigma$ level. Future facilities will perform significantly better. For example, the scalar solution \texttt{s7} would be compatible with Kerr at $1\sigma$ level with RXTE (cf.\ left panel of Fig.~\ref{fig:chi2_RXTE}) but it would be incompatible at more than $5\sigma$ level using eXTP (cf.\ top right panel of Fig.~\ref{fig:chi2}). In this case, an improvement by a factor of $5$ in the effective area of the detector is crucial.

\subsection{Tests with multiple QPO detections}
Let us now discuss the putative case in which more than two QPO triplets are detected from the same BH.
We follow the approach discussed in Refs.~\cite{Motta:2013wga,Maselli:2014fca}.
In particular, for each solution listed in Table~\ref{tab:parameters}, we choose $N$ equally spaced values of $\bar r$
in the range $\bar r=\left[1.1- 1.4\right]r_{\tn{ISCO}}$, and compute the corresponding set of $N$ QPO triplets
$(\nu_\varphi,\nu_\tn{nod},\nu_\tn{per})_{j=1,\dots,N}$.
We then express the azimuthal frequency $\nu_{\varphi}$ in terms of the radius by solving Eq.~\eqref{nu1GR} for $r$ (recall that the coordinate $R$ is related to $r$ by a simple shift). Upon replacing this expression into Eqs.~\eqref{nu2GR}--\eqref{nu3GR}, we obtain the functions $\nu^\tn{Kerr}_\tn{nod}(\nu_\varphi,M,a^\star)$ and $\nu^\tn{Kerr}_\tn{per}(\nu_\varphi,M,a^\star)$.
For each triplet, we simulate the QPO frequencies by drawing
from a Gaussian distribution centered at each given frequency $\nu_\varphi$ and with the same standard deviation as discussed before, namely we assume a relative error of $0.2\%$ on each frequency. For each drawn value of $\nu_\varphi$,
we span the $M\mu-a^\star$ parameter space and minimize the variable
\begin{equation}\label{chisquare}
\chi^2=\sum_{j=1}^{N}\left[\frac{(\nu_\tn{nod}-\nu^\tn{Kerr}_{\tn{nod}})_{j}^2}{\sigma^2_\tn{nod}}
+\frac{(\nu_\tn{per}-\nu^\tn{Kerr}_{\tn{per}})_{j}^2}{\sigma^2_\tn{per}}\right]\,,
\end{equation}
where $\nu_\tn{nod}$ and $\nu_\tn{per}$ are those drawn previously.
This procedure is equivalent to fitting the dependence of the simulated values
$\nu_\tn{nod}$ and $\nu_\tn{per}$ on $\nu_\varphi$ based on the assumptions that they were generated in the Kerr case.
The $\chi^2$  has average $E[\chi^2]=2N-2$ and standard deviation $\sigma[\chi^2]=2\sqrt{N-1}$.
Finally, the minimum of Eq.~\eqref{chisquare} corresponds to the best-fit
parameters $(\hat{M},\hat{a}^\star)$.

\begin{figure*}[ht]
\centering
\includegraphics[width=0.49\textwidth]{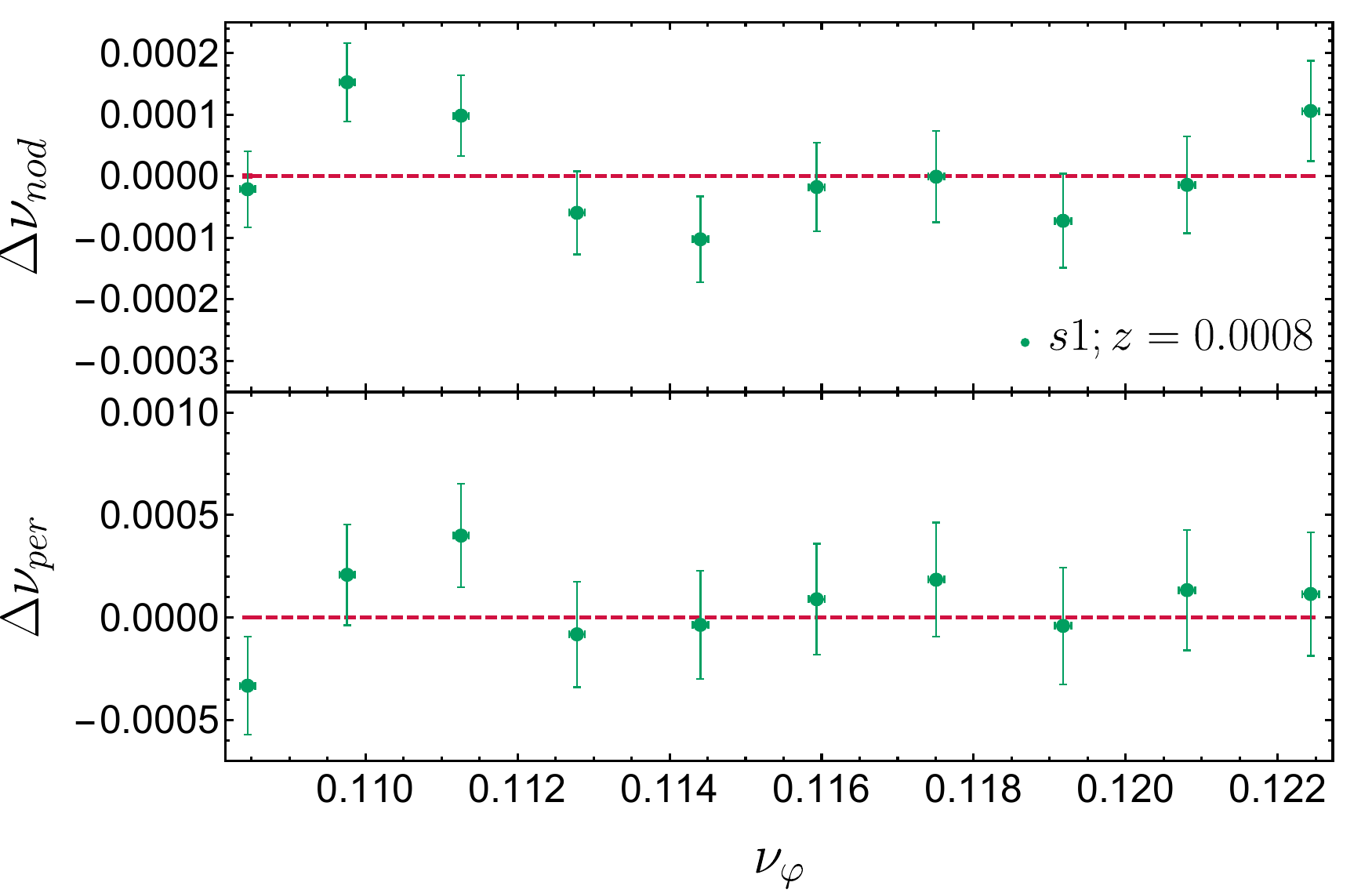}
\includegraphics[width=0.49\textwidth]{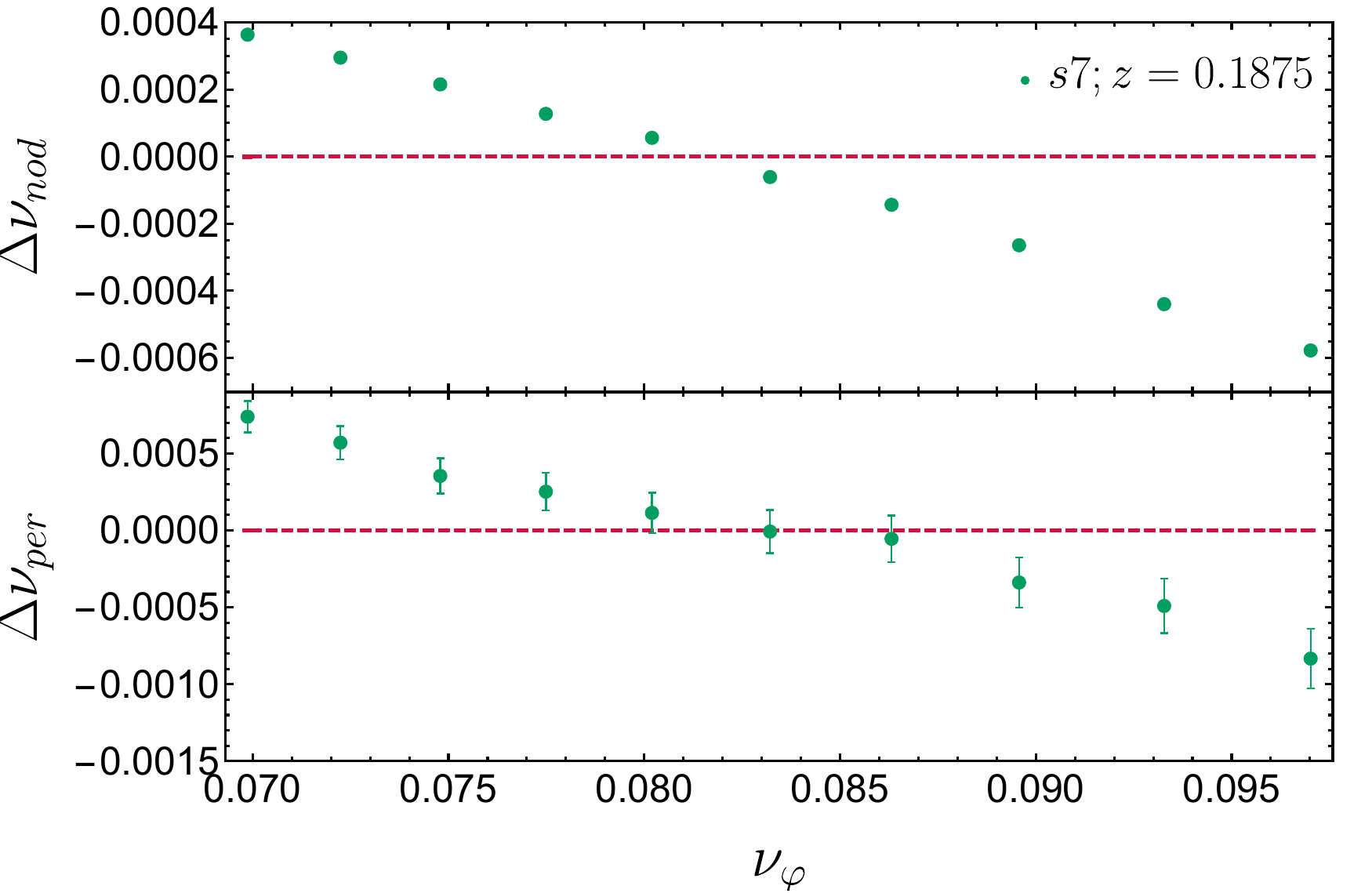}\\
\caption{The differences $\Delta\nu_\tn{nod}=(\nu_\tn{nod}-\hat{\nu}_\tn{nod})$
and $\Delta\nu_\tn{per}=(\nu_\tn{per}-\hat{\nu}_\tn{per})$ as functions of the azimuthal frequency
$\nu_\varphi$, for HBH solutions \texttt{s1} and \texttt{s7} listed in Table~\ref{tab:parameters}. The quantities $\hat{\nu}_\tn{per,nod}$ are the periastron and nodal frequencies computed for the best-fit parameters
$(\hat{M}\mu,\hat{a}^\star)$, which have been obtained minimizing the chi-square variable~\eqref{chisquare}. The error bar is $1\sigma$ and it is not visible on the scale of the top right panel. }
\label{fig:chisquareN}
\end{figure*}

As a representative example, in Fig.~\ref{fig:chisquareN} we show the results of this analysis for scalar HBH solutions \texttt{s1} and \texttt{s7} and considering $N=10$ multiple QPO triplets.
The former solution represents the case of our sample which is closer to Kerr ($q\approx 0.014$, $z\approx 0.0008$), whereas the latter solution could be already excluded by a two-triplet detection (cf.\ top right panel of Fig.~\ref{fig:chi2}).
In these cases, the minimization yields $\chi^2\simeq 20$ and  $\chi^2\simeq 8674$, respectively.
For the first dataset the $\chi^2$ is compatible with the expectation value $E[\chi^2]=18$ within  $2\sigma$, while for \texttt{s7}
the $\chi^2$ is incompatible at more than $1400\sigma$.
This already indicates that the observed data would disagree with the theoretical model based on the assumption of a Kerr BH.
Furthermore, for \texttt{s1} the values of $\Delta\nu_\tn{per}=(\nu_\tn{per}-\hat{\nu}_\tn{per})$ and $\Delta\nu_\tn{nod}=(\nu_\tn{nod}-\hat{\nu}_\tn{nod})$
are all distributed around zero roughly within $1\sigma$, while for \texttt{s7} the plot shows a clear trend, identifying a correlation with the azimuthal frequency.

\subsection{QPO constraints on boson and Proca stars}\label{sec:analysisBS}

Besides allowing for Kerr BHs with bosonic hair, model~\eqref{theory}
predicts the existence of self-gravitating solitons
(i.e. smooth particle-like configurations without an event horizon)
sustained by the wave-like, dispersive nature of the bosonic field.
In the scalar and vector case these solutions are called boson stars and Proca stars, respectively (for a review on boson stars, see~\cite{Liebling:2012fv}).
The same techniques discussed above can be applied to boson- and Proca-star configurations, in order to test whether the X-ray signal from these objects can be used to distinguish them from a Kerr BH, or to simply detect them.

The geodesic structure of spinning boson/Proca stars is remarkably different from that of a Kerr BH --~see, e.g., Refs.~\cite{Grandclement:2014msa,Meliani:2015zta} for earlier studies. These solutions can be sorted by their position along the red curve in the left panels of Fig.~\ref{fig:parameterspace} in a counter-clockwise sense\footnote{Note that, for the branch located before the minimum frequency, the solutions can be equivalently sorted by increasing values of the parameter $z$, which in the case of scalar/Proca stars simply reads $z:=1-w^2/\mu^2$, cf.~Eq.~\eqref{zeta},
with $0<z<0.584$ for boson stars and $0<z<0.445$ for Proca stars.}, with the first solutions having the smallest compactness~\cite{Herdeiro:2015gia}. Because the bosonic field extends from the center of the star to infinity, a boson/Proca star does not have a well defined radius, at variance with the case of perfect-fluid stars. An effective size can be defined by the radius $R_{99}$ that contains $99\%$ of the bosonic mass~\cite{Liebling:2012fv,Herdeiro:2015gia}, cf.~Table~\ref{tab:parameters}. Nonrelativistic solutions (i.e., those with $R_{99}\gg M$), clearly cannot model compact objects. We focus here on relativistic solutions, which lie near the maximum of the red curve in the left panels of Fig.~\ref{fig:parameterspace}.

\begin{figure*}[ht]
\centering
\includegraphics[width=\textwidth]{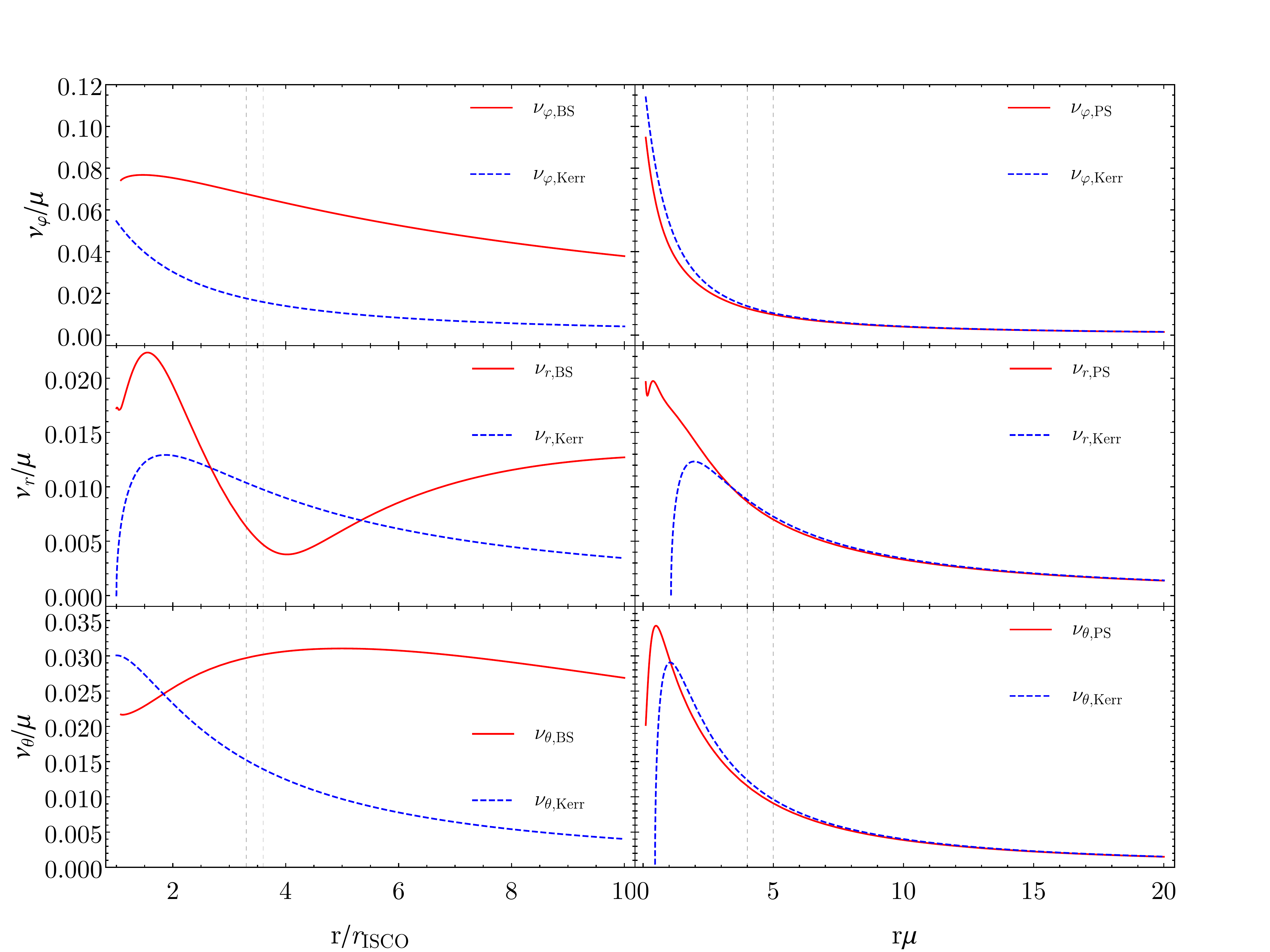}
\caption{Same as in Fig.~\ref{fig:frequencies} but for the boson-star solution \texttt{s15}$^\star$ (left panels) and for the Proca star solution \texttt{v15}$^\star$ (right panels). The vertical dashed lines denote the radii that we have considered in the $\chi^2$ analysis of Fig.~\ref{fig:chi2BS}.}
\label{fig:BSs} \end{figure*}

Relativistic boson stars display qualitative differences relative to the Kerr case.
For example, solutions \texttt{s13}$^\star$-\texttt{s15}$^\star$ admit circular co-rotating orbits only for $\bar{r}$ larger than a critical radius, below which the angular frequency in Eq.~\eqref{Omega} becomes complex. These orbits are stable so the critical radius is effectively the ISCO of these geometries, even though in this case $\nu_r$ at the ISCO is nonzero.
As show in Fig.~\ref{fig:BSs}, the epicyclic frequencies near the ISCO of these objects are completely different from the corresponding Kerr case with the same mass and spin.

\begin{figure*}[ht] \centering
\includegraphics[width=0.48\textwidth]{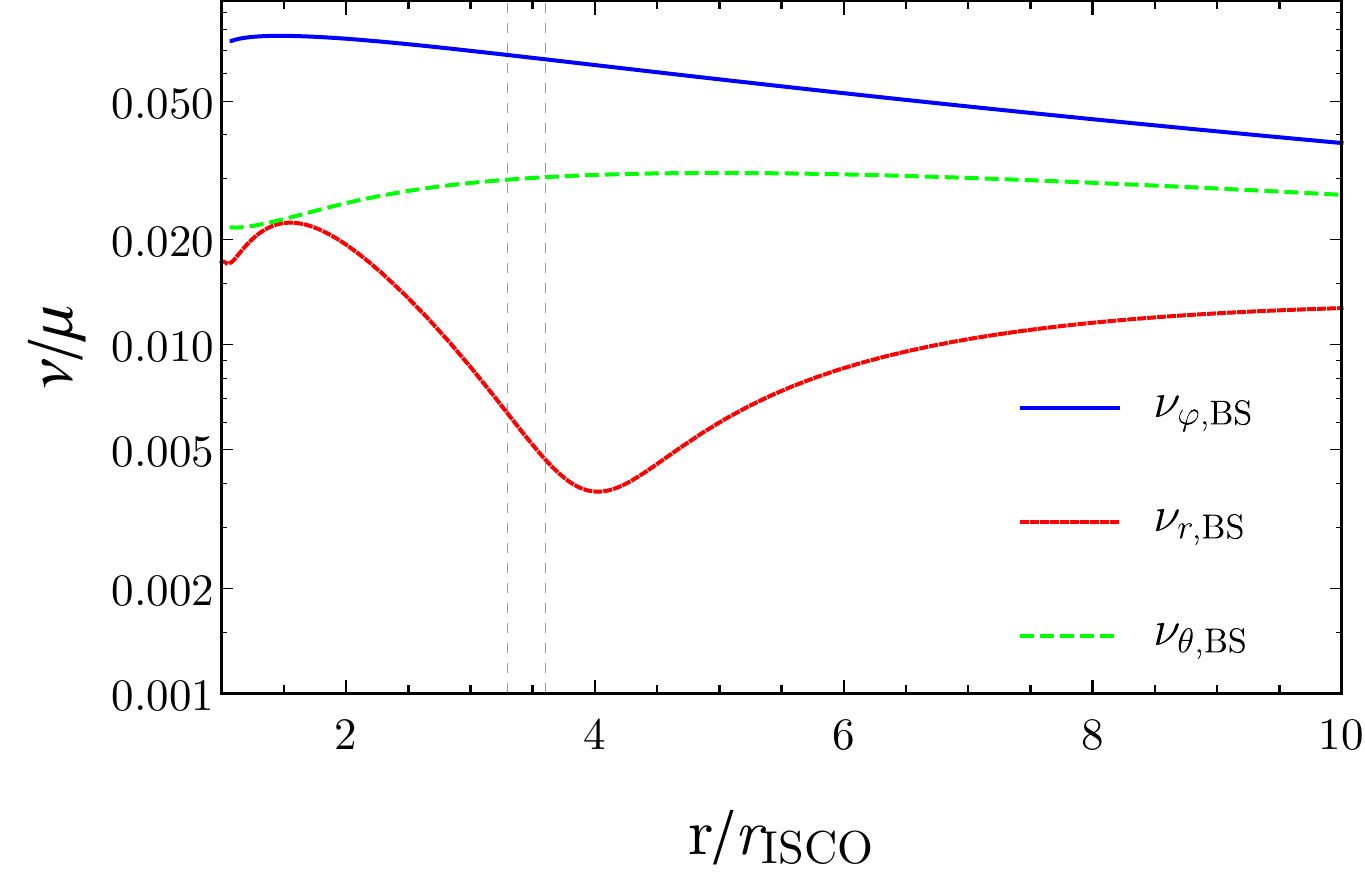}
\includegraphics[width=0.48\textwidth]{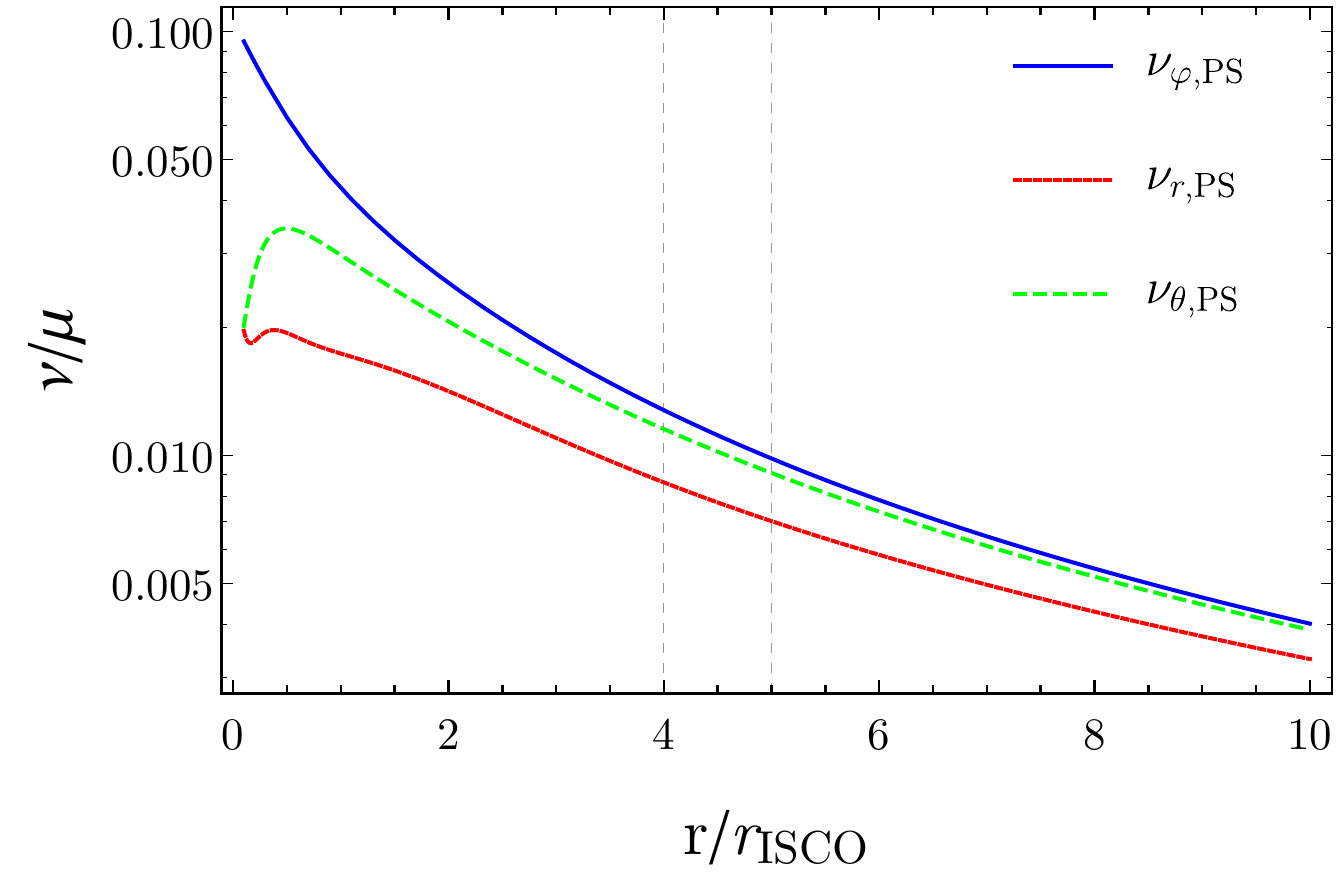}
\caption{Same as Fig.~\ref{fig:BSs} but showing the azimuthal and epicyclic frequencies in the same plot.
\label{fig:allfreq_BS}}
\end{figure*}

Indeed, the same analysis performed in the BH case can even provide a null result for boson stars. If the emission radius of the QPO triplet $(\nu_\varphi,\nu_{\tn{per}},\nu_{\tn{nod}})$ is near the ISCO, either the inversion to obtain the mass, spin and emission radius of a Kerr BH is impossible, or such inversion gives a value of the Kerr spin exceeding the Kerr bound\footnote{
The Kerr bound is exceeded by a subset of configurations between the vacuum limit ($M=J=0$) and $0<z<0.149$ for  boson stars or $0<z<0.0736$ for Proca stars.}, i.e.~$a^\star>1$.
In principle, there might exist a different triplet (e.g., a linear combination of $\nu_\varphi$, $\nu_{\tn{per}}$, $\nu_{\tn{nod}}$) which yields an invertible relation. However, such new combination of epicyclic frequencies should have an intrinsic geometrical meaning and should therefore be valid for any object (including Kerr BHs) and for any emission radius. Given the dramatic differences in the epicyclic frequencies shown in Fig.~\ref{fig:BSs}, it is very unlikely that a combination thereof would be similar to the Kerr counterpart for all boson/Proca stars. 
On the other hand, the standard inversion is not problematic if the emission occurs at larger distances, since for larger values of $\bar{r}$ the frequencies approach their asymptotic values, independently of the structure of the central object [cf.\ Figs.~\ref{fig:BSs} and \ref{fig:allfreq_BS}].

\begin{figure*}[ht]
\centering
\includegraphics[width=0.42\textwidth]{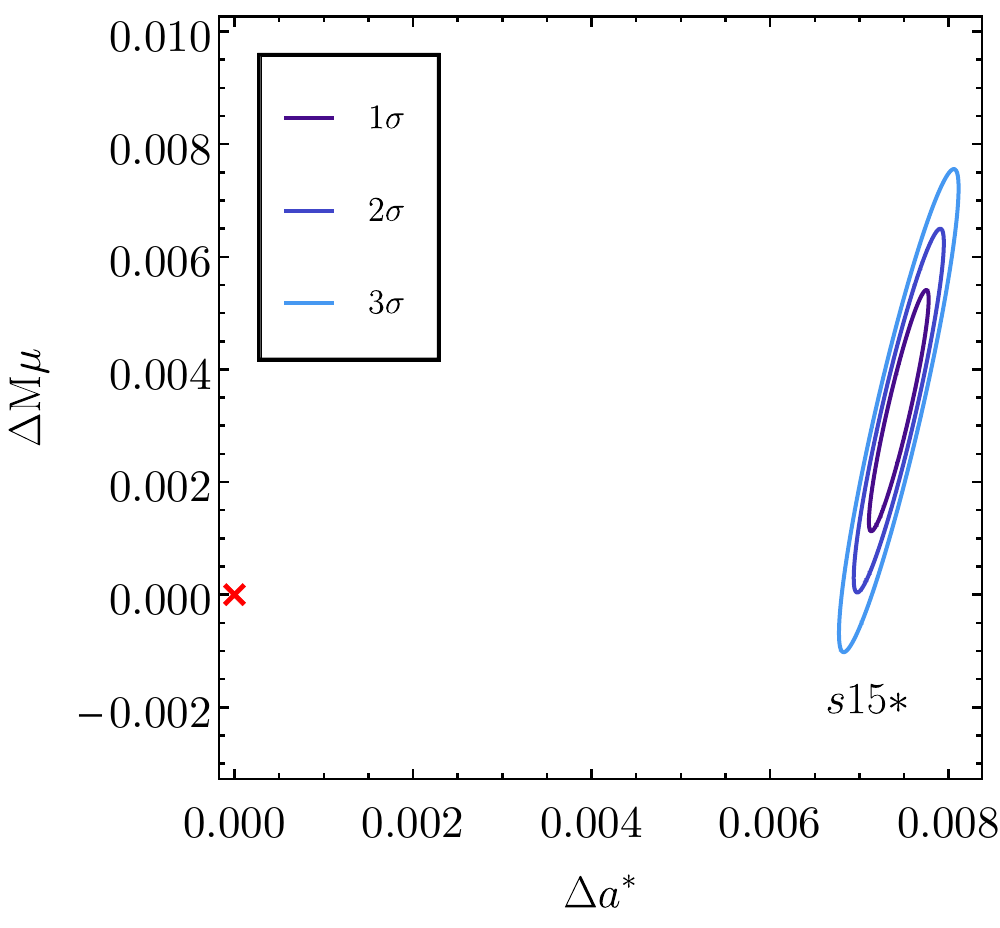}
\includegraphics[width=0.425\textwidth]{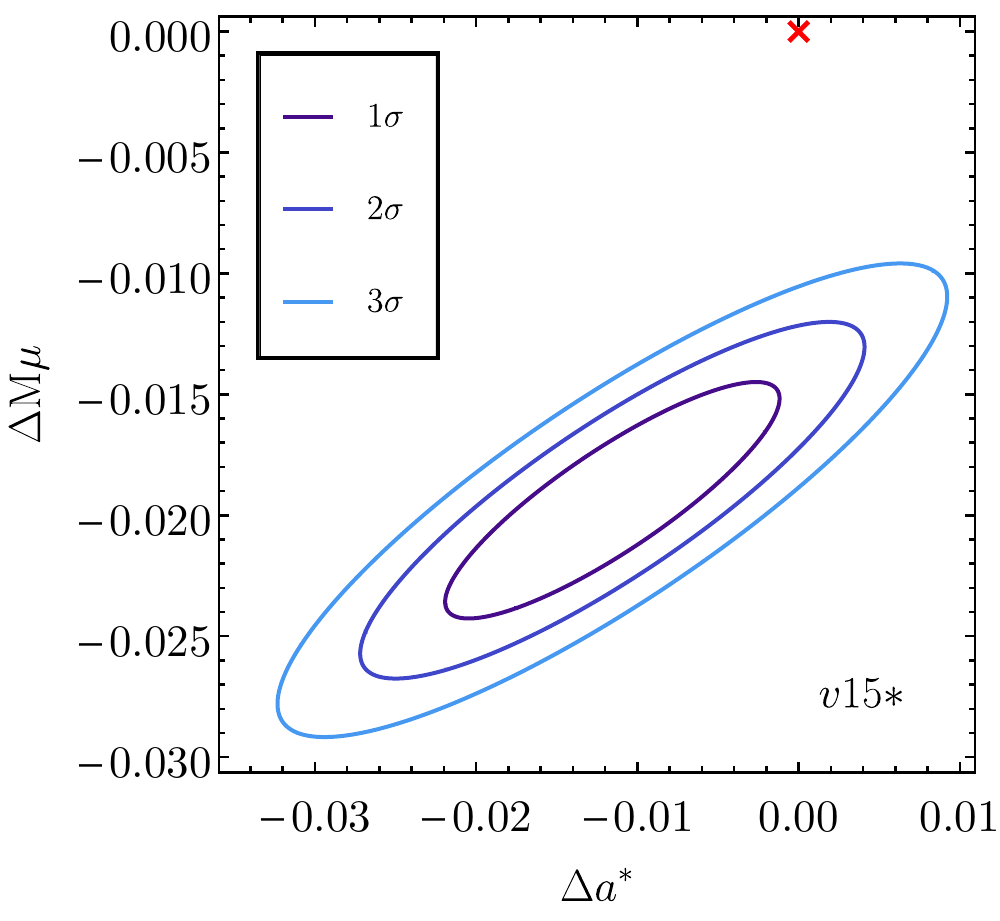}
\caption{Same as in Fig.~\ref{fig:chi2} but for the same boson-star solution \texttt{s15}$^\star$ (left panel) and for the Proca star solution \texttt{v15}$^\star$ (right panel), whose geodesic frequencies are shown in Fig.~\ref{fig:BSs}.
}
\label{fig:chi2BS} \end{figure*}

As a representative example, for solution \texttt{s15}$^*$ the inversion gives $a^\star>1$ when the emission radius $\bar{r}\lesssim 3.2 \,r_{\rm ISCO}$. This already suggests that it is very unlikely to mistake the QPO emission of an accreting Kerr BH by that of an accreting boson star within the RPM, since it is natural to expect that the QPO signal originates deep inside the gravitational potential of the object and near the ISCO, where the density of the accretion disk peaks.
Furthermore, even in the unlikely case in which the emission happens at larger distances, the geodesic structure of boson/Proca stars might be so different from that of a Kerr BH that it would be easy to distinguish between the two cases. A representative example is shown in the left panel of Fig.~\ref{fig:chi2BS}, where we apply the $\chi^2$ analysis previously presented to solution \texttt{s15}$^*$, in which we consider two emission radii at $3.3\,r_{\rm ISCO}$ and $3.6\,r_{\rm ISCO}$, which are slightly larger than the critical radius below which the inversion of the QPO triplet is meaningless.

The situation for Proca stars is even more dramatic, since the solutions listed in Table~\ref{tab:parameters} do not have an ISCO for corotating orbits.  (This property holds true also for other Proca stars that we have analyzed but which were not reported in Table~\ref{tab:parameters}.) Therefore, it is hard to conceive a mechanism that would produce a cut-off frequency associated to the accretion flow for these objects. Matter on the equatorial plane would simply inspiral down all the way to the center of the star or --~in case of some coupling to the vector field~-- to some effective ``surface'', where the vector-field density is nonnegligible (we recall that boson/Proca stars do not have a definite surface).  In both cases, the emission is expected to be completely different from that of an accretion disk which extends down to the ISCO.
Nonetheless, for completeness, in the right panels of Figs.~\ref{fig:BSs} and \ref{fig:allfreq_BS} we show the azimuthal and epicyclic frequencies for the Proca-star solution \texttt{v15}$^*$, whereas in the right panel of Fig.~\ref{fig:chi2BS} we show the $\chi^2$ analysis for the same object, in which we consider two emission radii at $4.76 M$ and at $5.95 M$.
Note that the emission radius has some degree of arbitrariness in this case, due to the absence of an ISCO that would set the scale of an accretion disk. As shown in the right panels of Fig.~\ref{fig:BSs}, at large emission radii the frequencies are similar to their Kerr counterpart. However, the relative deviations are at the level of a few percent, i.e. larger than the observational errors of future detectors. This explains why, even if we consider relatively large emission radii, the $\chi^2$ analysis shown in the right panel of Fig.~\ref{fig:chi2BS} excludes this solution at more than $3\sigma$ level. Had we chosen smaller emission radii, such deviations would have been even larger.

We have applied the above analysis to several boson/Proca star solutions with various emission radii, finding results that are always qualitatively similar to those shown in Fig.~\ref{fig:chi2BS}; this suggests that boson/Proca star solutions would be easily distinguishable from a Kerr BH using QPO diagnostic.

\section{Discussion and conclusion}\label{sec:concl}

Future large-area X-ray detectors, such as the upcoming eXTP, will allow for precision QPO spectroscopy of accreting compact objects.
In this work, we have shown that, if supported by a solid astrophysical model, the QPO spectrum can be used to distinguish between BHs with bosonic hair and Kerr BHs. The former are stationary solutions of standard general relativity minimally coupled to a (complex) massive scalar or massive vector field. Furthermore, we have shown that the same technique can be straightforwardly applied to boson/Proca stars which are also predicted by Einstein's theory in the presence of light bosonic fields.

We have performed a thorough geodesic analysis on several numerical solutions for spinning BHs and boson/Proca stars with arbitrary (i.e.\ not necessarily small) spin, which have been recently obtained in Refs.~\cite{Herdeiro:2014goa,Herdeiro:2015gia,Herdeiro:2016tmi}.
We have identified a phenomenological parameter $z$, which is a combination of the extra Noether charge of the solutions [cf. Eq.~\eqref{zeta}], and which provides a good indicator of the deviations from the Kerr metric.

In particular, our main results can be summarized as follows:
\begin{itemize}
 \item Assuming the RPM as the model to explain the QPO frequencies, our analysis suggests that BHs with scalar (resp. vector) hair with $z\gtrsim 0.12$ (resp. $z\gtrsim 0.025$) can be excluded by future instruments at more than $3\sigma$ level (assuming, of course, the observational data will be compatible with Kerr).
 \item The geodesic structure of  boson and Proca stars is so peculiar that their QPO spectrum might even be incompatible with the RPM applied to the Kerr case. For these systems, the possibility of inverting the expressions of the QPO triplet to obtain the properties of a corresponding Kerr BH depends on the emission radius and would in general give inconsistent results if the emission radius is very close to the ISCO, as expected for realistic accretion-disk models.
 \item All Proca stars that we have analyzed do not even possess an ISCO; this would drastically affect the dynamics of accretion and, in turn, the X-ray flux.
\item Our results suggest that the QPO triplet~\eqref{freq} observed in the X-ray spectrum of GRO J1655-40~\cite{2005ApJ...629..403C,Motta:2013wga} cannot be generated by either a boson or a Proca star, at least if the nature of the QPOs is predominantly geodesic. Indeed, such frequencies would either be incompatible to the RPM (i.e., they would not correspond to a single mass, spin, and emission radius) or they would correspond to an emission radius much larger than the ISCO (if the latter exist), and therefore highly disfavored by astrophysical considerations. Furthermore, in the latter case the physical parameters (mass and spin) of the object extrapolated from the QPO inversion could be largely inconsistent with the expectations or with independent astrophysical observations (we recall that the mass of GRO J1655-40 inferred from optical/NIR spectro-photometric observations is $(5.4\, \pm 0.3) M_\odot$~\cite{Beer_Podsiadlowski_2002}). Thus, under our working assumptions, the QPO detection from GRO J1655-40 rules out the possibility that this compact source is a boson or a Proca star.
\end{itemize}

To model the QPOs we have adopted a simple model (the RPM) as a proxy to interpret the QPOs in terms of geodesic frequencies, but most of the QPO models (including more realistic ones~\cite{1999PhR...311..259W,2003MNRAS.341..832Z,2004ApJ...617L..45B,2005AN....326..820K,2006CQGra..23.1689A,2015MNRAS.446..240M,2016MNRAS.456.3245M,2015arXiv151007414M}) assume some correlation between the QPO frequency and geodesic motion of matter in the accretion disk, and would provide similar results. For example, the constraints coming from QPO spectroscopy using the RPM or a modified Epicyclic Resonance model~\cite{Abramowicz:2004je} are similar~\cite{Maselli:2017kic}. This is expected since all these models predominantly depend on the geometry of the accretion disk and, in particular, on the ISCO location.

In order for the metric to be stationary, the bosonic field needs to be complex. However, light real bosonic fields might give rise to solutions which are dynamical only on extremely long time scales and which are presumably qualitatively similar to the stationary solutions presented here. This case has been studied for oscillatons~\cite{Seidel:1991zh} --~which are the weakly-dynamical counterparts of boson stars~-- and for BHs with scalar hair in the adiabatic approximation and neglecting backreaction~\cite{Brito:2014wla}. We expect that our analysis can be straightforwardly extended also to these cases, providing potentially interesting constraints on real bosonic fields (such as axions and dark photons) near compact objects.

Finally, our results confirm the enormous potential of QPO spectroscopy for strong-field tests of gravity (see also Refs.~\cite{Bambi:2013fea,Maselli:2014fca,Maselli:2017kic}). Because these tests are less limited by instrumental precision than by astrophysical systematics, we strongly advocate the development of reliable precise astrophysical models of QPO signal from accreting BHs.


\section*{Acknowledgments}
We are indebted to Luigi Stella for interesting discussions on different QPO models.
NF's research is supported in part by the European Research Council under the European Union's Seventh Framework Programme (FP7/2007-2013) / ERC grant agreement n. 306425 ``Challenging General Relativity''. PP acknowledges support from FCT-Portugal through project IF/00293/2013. C.H. and E.R. acknowledge support by the FCT-Portugal IF programme, by the CIDMA (FCT) strategic project UID/MAT/04106/2013. We also acknowledge support by the EU grants NRHEP-295189-FP7-PEOPLE-2011-IRSES and H2020-MSCA-RISE-2015 Grant No. StronGrHEP-690904, and by the COST Action CA16104.
\vskip 2cm


\bibliography{bibnote}

\end{document}